\documentclass[conference]{IEEEtran}
\IEEEoverridecommandlockouts
\usepackage{cite}
\usepackage{amsmath,amssymb,amsfonts,amsthm}
\usepackage{bbm}
\theoremstyle{plain}
\newtheorem{theorem}{\bf{Theorem}} 
\newtheorem{lemma}{\bf{Lemma}} 
\newtheorem{assumption}{\bf{Assumption}} 
\theoremstyle{remark}
\newtheorem{remark}{\bf{Remark}} 
\usepackage{algorithm} 
\usepackage{algorithmic}
\usepackage{makecell} 
\usepackage{graphicx}
\usepackage{subfigure}
\usepackage{textcomp}
\usepackage{xcolor}
\usepackage{bm}
\usepackage{multirow}
\usepackage{setspace}
\usepackage[normal]{caption}
\renewcommand{\raggedright}{\leftskip=0pt \rightskip=0pt plus 0cm} 
\usepackage{enumitem}
\setlist{leftmargin=1em}
\setitemize[1]{itemsep=0pt,partopsep=0pt,parsep=\parskip,topsep=0pt}
\usepackage{amssymb}  
\newcommand*{\QEDB}{\hfill\ensuremath{\square}} 
\DeclareMathOperator{\Tr}{Tr} 
\usepackage{hyperref}
\usepackage[hyphenbreaks]{breakurl}

\def\BibTeX{{\rm B\kern-.05em{\sc i\kern-.025em b}\kern-.08em
    T\kern-.1667em\lower.7ex\hbox{E}\kern-.125emX}}
\begin{document}

\title{Semi-Decentralized Federated Edge Learning for Fast Convergence on Non-IID Data}
\author{\IEEEauthorblockN{Yuchang Sun$^*$, Jiawei Shao$^*$, Yuyi Mao$^\dag$, Jessie Hui Wang$^\ddag$, and Jun Zhang$^*$}
\IEEEauthorblockA{$^*$Dept. of ECE, The Hong Kong University of Science and Technology, Hong Kong\\
$^\dag$Dept. of EIE, The Hong Kong Polytechnic University, Hong Kong \\
$^\ddag$Institute for Network Sciences and Cyberspace, BNRist, Tsinghua University, Beijing, China \\
Email: \{yuchang.sun, jiawei.shao\}@connect.ust.hk,
yuyi-eie.mao@polyu.edu.hk, \\
jessiewang@tsinghua.edu.cn,
eejzhang@ust.hk}
}
\maketitle

\begin{abstract}
Federated edge learning (FEEL) has emerged as an effective approach to reduce the large communication latency in Cloud-based machine learning solutions, while preserving data privacy. Unfortunately, the learning performance of FEEL may be compromised due to limited training data in a single edge cluster. In this paper, we investigate a novel framework of FEEL, namely \textit{semi-decentralized federated edge learning} (SD-FEEL). By allowing model aggregation across different edge clusters, SD-FEEL enjoys the benefit of FEEL in reducing the training latency, while improving the learning performance by accessing richer training data from multiple edge clusters. A training algorithm for SD-FEEL with three main procedures in each round is presented, including local model updates, intra-cluster and inter-cluster model aggregations, which is proved to converge on non-independent and identically distributed (non-IID) data. We also characterize the interplay between the network topology of the edge servers and the communication overhead of inter-cluster model aggregation on the training performance. Experiment results corroborate our analysis and demonstrate the effectiveness of SD-FFEL in achieving faster convergence than traditional federated learning architectures. Besides, guidelines on choosing critical hyper-parameters of the training algorithm are also provided.
\end{abstract}


\section{Introduction}
The recent development of machine learning (ML) technologies has led to major breakthroughs in various domains. Meanwhile, the number of Internet of Things (IoT) devices is growing at a drastic speed, and it is envisaged to reach more than 30 billion IoT devices by 2025 \cite{IoT}. This results in a huge volume of data generated at the edge of the wireless network, which will facilitate the training of powerful ML models. A traditional approach is to upload these data to a centralized server for model training. However, offloading data raises severe privacy concerns as data collected by IoT devices, e.g., smartphones and healthcare sensors, may contain privacy-sensitive information \cite{meneghello2019iot}.

To resolve the privacy issues in centralized ML, federated learning (FL) was proposed by Google in 2017 as a privacy-preserving distributed ML paradigm \cite{mcmahan2017communication}. FL enables IoT devices to collaboratively learn a shared ML model without disclosing their local data. A typical FL system consists of a Cloud-based parameter server (PS) and a number of client nodes which train a deep learning (DL) model via multiple global iterations. In each global iteration, the client nodes first download the global model maintained by the PS, based on which several rounds of local updates are performed using the private local data. The updated local models are then transmitted to the PS for global model aggregation. Nevertheless, as DL models, typically with millions of parameters, need to be uploaded to the Cloud in every global iteration, the communication latency caused by core-network congestion significantly bottlenecks the learning efficiency \cite{li2020federated}, \cite{shi2020communication}.

Attributed to the emergence of mobile edge computing (MEC) \cite{mao2017survey}, federated edge learning (FEEL), where an edge server located in close proximity to the client nodes (e.g., a base station) is deployed as the PS, was proposed as a promising alternative to Cloud-based FL \cite{lim2020federated}. Despite its great promise in reducing model uploading latency, one of the major challenges in FEEL systems is the unstable wireless connections between the client nodes and the edge server, and client nodes with unfavorable channel conditions may become stragglers for model aggregation and in turn slow down the training process. As a result, improving the training efficiency of FEEL has drawn great interests in existing literature. Considering the limited resources, a control algorithm to determine the global model aggregation frequency of FEEL was developed in \cite{wang2019adaptive} to maximize the learning performance. In order to reduce the training time, a client selection algorithm was proposed for FEEL in \cite{nishio2019client}, which eliminates the straggling client nodes from global model aggregation. Besides, gradient quantization and sparsification techniques were also adopted to achieve communication-efficient FEEL \cite{mills2019communication}, \cite{amiri2020federated}. However, the number of accessible client nodes of a single edge server is generally limited, which hinders the benefits of FL due to the small amount of training data.

To unleash the full potential of FEEL, recent works exploited cooperation among multiple edge servers for model training. In \cite{liu2020client}, a client-edge-cloud hierarchical FL system along with a training algorithm, namely HierFAVG, was developed, which takes advantages of the Cloud and edge servers to accelerate model training. Similar investigations in \cite{wang2020local} extended HierFAVG to a multi-level stochastic gradient descent (SGD) algorithm. Besides, a fog-assisted FL framework was proposed in \cite{saha2020fogfl}, where the Cloud-based PS selects an optimal fog node to serve as the global model aggregator in each communication round. Unfortunately, these works still rely on the connection to the Cloud, which may suffer from excessive communication latency. Besides, such architectures are vulnerable to the single-point failure and not scalable to large-scale deployment.

In this paper, we investigate a novel FL architecture, namely semi-decentralized federated edge learning (SD-FEEL), to improve the training efficiency. This architecture is motivated by the low communication latency between edge servers so that efficient model exchanges can be realized. Specifically, we consider multiple edge servers, and each of them coordinates a cluster of client nodes to perform local model updating and \textit{intra-cluster} model aggregation. The edge servers periodically share their updated models to the neighboring edge servers for \textit{inter-cluster} model aggregation. Such a semi-decentralized training protocol can accelerate the learning process since a large number of client nodes can collaborate with low communication cost. It is worthy mentioning that a similar design was investigated in \cite{castiglia2020multi}, which, however, was limited to an idealized independent and identically distributed (IID) assumption of the local training data. Our work in this paper advances SD-FEEL in the following aspects:
\begin{itemize}[leftmargin=1em]
    \item We extend SD-FEEL to the case with non-IID data. To investigate impacts of the network topology on learning performance, the edge servers are allowed to share and aggregate models multiple times in each round of inter-cluster model aggregation.
    \item Convergence of a training algorithm supporting SD-FEEL on non-IID data is proved. Our analysis shows when both intra-cluster and inter-cluster model aggregations are performed at each iteration, SD-FEEL reduces to the vanilla SGD \cite{parallelsgd} given mutually connected edge servers. This is also achievable by multiple times of inter-server model sharings when the edge servers are only partially connected. 
    \item Extensive experiments are conducted to demonstrate the benefits of SD-FEEL in achieving faster convergence compared to existing FL schemes. Our results also provide insightful guidelines on choosing the intra-/inter-cluster model aggregation frequencies and the communication strategy among edge servers to reduce the training latency.
\end{itemize}

The rest of the paper is organized as follows. Section \ref{sec-2} introduces the framework of semi-decentralized FEEL. Section \ref{sec-3} shows the convergence of SD-FEEL on non-IID data and discusses various insights. We present experiment results in Section \ref{sec-4} and conclude this paper in Section \ref{sec-5}.

\section{Semi-Decentralized FEEL}\label{sec-2}
In this section, we first introduce the semi-decentralized FEEL system, and a training algorithm is then presented.
\subsection{Semi-Decentralized FEEL System}
\begin{figure}[t]
    \centering{
    \includegraphics[width=0.9\linewidth]{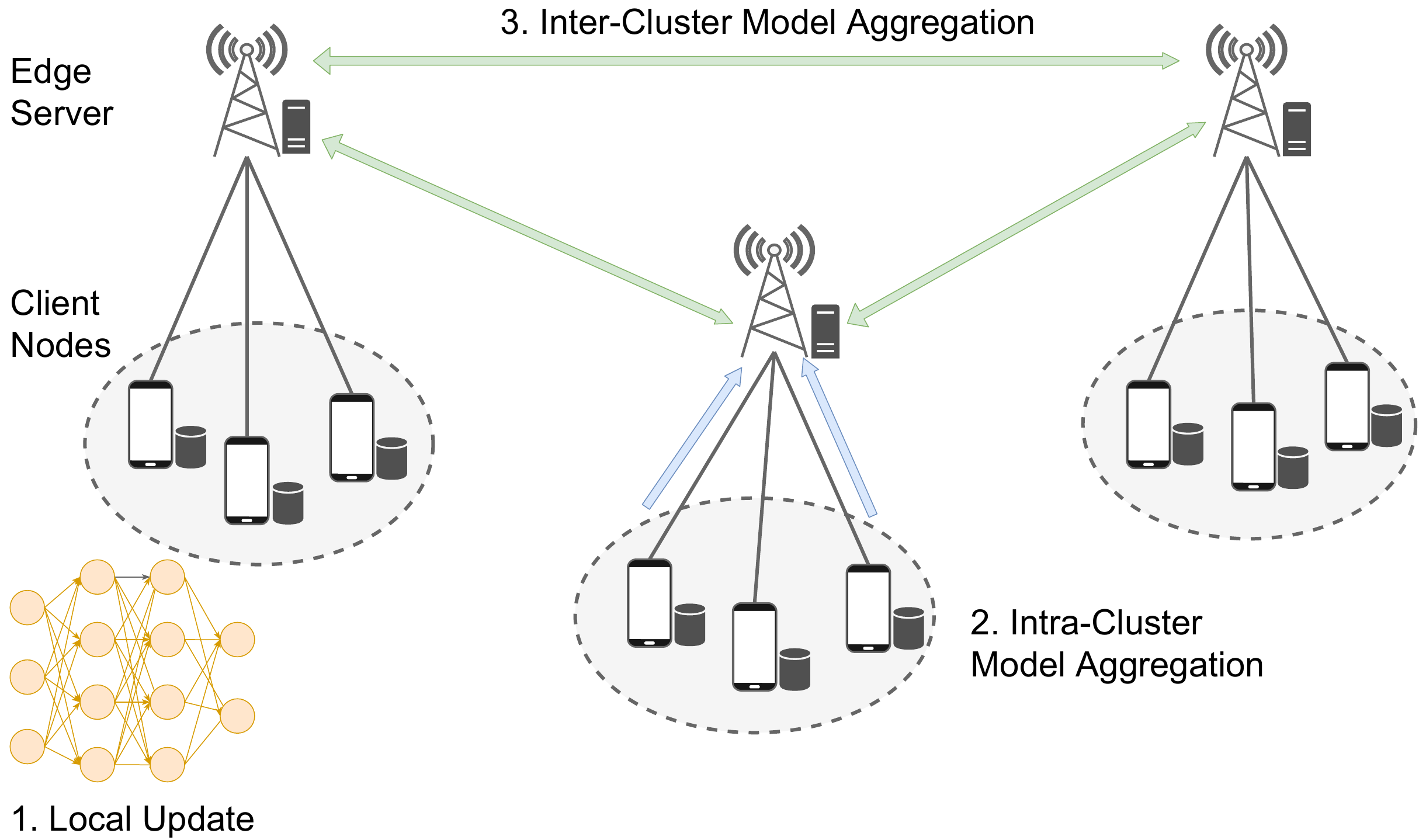}}
    \caption{The semi-decentralized FEEL system.}
    \label{fig-flow}
\end{figure}

We consider an edge-assisted federated learning system as shown in Fig. \ref{fig-flow}, which consists of $C$ client nodes (denoted as set $\mathcal{C}$) and $D$ edge servers (denoted as set $\mathcal{D}$). Each client node is associated with an edge server according to some pre-defined criteria, e.g., physical proximity and network coverage. We denote the set of client nodes associated with the $d$-th edge server as $\mathcal{C}_{d}$, which form one edge cluster. Some of the edge servers are inter-connected via high-speed cables, which facilitates information exchange in SD-FEEL. We use a connected graph $\mathcal{G}$ to model the connectivity among the edge servers, and denote the one-hop neighbors of the $d$-th edge server as $\mathcal{N}_d$.

Each client node is assumed to have a set of local training data, denoted as $\mathcal{S}_i = \{\bm{x}_{j},y_{j}\}_{j=1}^{|\mathcal{S}_i|}, i\in\mathcal{C}$, where $\bm{x}_{j}$ is the data sample and $y_{j}$ is the corresponding label. The whole training dataset and the collection of local training data across the client nodes associated with the $d$-th server are denoted as $\mathcal{S}$ and $\Tilde{\mathcal{S}}_d$, respectively. The client nodes collaborate to train a shared DL model, denoted as $\bm{w}\in\mathbb{R}^{M}$ with $M$ as the number of trainable parameters. Let $f(\bm{x}_j,y_j; \bm{w})$ be the loss function associated with the data sample $(\bm{x}_j,y_j)$ based on the model $\bm{w}$. As a result, the objective of FL is to minimize the loss function over all the training data, i.e.,
\begin{equation}
    \min_{\bm{w}\in\mathbb{R}^{M}} \left\{ F(\bm{w})\triangleq \sum_{i\in\mathcal{C}} \frac{|\mathcal{S}_i|}{|\mathcal{S}|} F_i(\bm{w}) \right\},
    \label{gl-func}
\end{equation}
where $F_i(\bm{w}) \triangleq \frac{1}{|\mathcal{S}_i|} \sum_{j\in\mathcal{S}_i} f(\bm{x}_j, y_j; \bm{w})$.

It is important to note that we will not pose the assumption of homogeneous (i.e., IID) data across the client nodes, which is more reflective of real scenarios and also a highly non-trivial extension compared to the existing study \cite{castiglia2020multi}.
\subsection{Training Algorithm}
Without loss of generality, we assume there are $K$ iterations in the training process, which contains three main procedures, including 1) local model update, 2) intra-cluster model aggregation, and 3) inter-cluster model aggregation. While local model updates are performed at the client nodes, the edge servers are in charge of the other two procedures. The periods of intra-cluster and inter-cluster model aggregation, denoted as $\tau_1$ and  $\tau_2$ respectively, are two critical hyper-parameters in the training algorithm.
\subsubsection{Local Model Update} In the $k$-th iteration, each client node performs model update with its local data using the mini-batch SGD algorithm according to the following expression:
\begin{equation}
    \bm{w}_{k}^{(i)} \leftarrow \bm{w}_{k-1}^{(i)} - \eta g(\bm{\xi}_{k}^{(i)}; \bm{w}_{k-1}^{(i)}), i\in\mathcal{C},
    \label{sgd}
\end{equation}
where $\bm{w}_{k-1}^{(i)}$ is the most updated model available at the $i$-th client node at the start of the $k$-th iteration, $g(\bm{\xi}_{k}^{(i)}; \bm{w}_{k-1}^{(i)})$ is the gradient computed on the batch of randomly-sampled local training data $\bm{\xi}_{k}^{(i)}$, and $\eta$ denotes the learning rate.
\addtolength{\topmargin}{0.01in}
\subsubsection{Intra-cluster Model Aggregation} In the $k$-th iteration (when $k$ is a multiple integer of $\tau_1$ but not $\tau_1\tau_2$), each client node uploads its most updated model to the associated edge server, and the edge server aggregates the received local models according to the following expression:
\begin{equation}
    \Tilde{\bm{w}}_{k}^{(d)} \leftarrow \sum_{i\in \mathcal{C}_d} \frac{\left|\mathcal{S}_i\right|}{\left| \Tilde{\mathcal{S}}_d \right|} \bm{w}_{k}^{(i)}, d\in\mathcal{D},
    \label{intra}
\end{equation}
where $\Tilde{\bm{w}}_{k}^{(d)}$ denotes the intra-cluster aggregated model at the $d$-th edge server. After that, $\Tilde{\bm{w}}_{k}^{(d)}$ is broadcasted to the client nodes $\mathcal{C}_d$, which can be expressed as follows: 
\begin{equation}
    \bm{w}_{k}^{(i)} \leftarrow \Tilde{\bm{w}}_{k}^{(d)}, i\in\mathcal{C}_d.
    \label{distribute}
\end{equation}
\subsubsection{Inter-cluster Model Aggregation} In the $k$-th iteration (when $k$ is a multiple integer of $\tau_1\tau_2$), after intra-cluster model aggregation, each edge server also shares its intra-cluster aggregated model to the one-hop neighbouring edge servers for inter-cluster model aggregation, before sending it back to the client nodes. Each round of inter-cluster model aggregation consists $\alpha \!\in\! \{\!1,\!2,\!\dots\!\}$ times of model sharing and aggregation, which can be expressed as follows:
\begin{equation}
    \hat{\bm{w}}_{k,l}^{(d)} \leftarrow \sum_{j\in \mathcal{N}_d\cup\{d\}} p_{j,d} \hat{\bm{w}}_{k,l-1}^{(j)}, l=1,2,\dots,\alpha,
    \label{inter}
\end{equation}
where $\hat{\bm{w}}_{k,0}^{(d)}=\Tilde{\bm{w}}_{k}^{(d)}$ is the intra-cluster aggregated model at the $d$-th edge server, and $\mathbf{P}\!\triangleq\![p_{i,j}]\in \mathbb{R}^{D\!\times\!D}$ is termed as the mixing matrix. We choose $\mathbf{P}= \mathbf{I}-\frac{2}{\lambda_1(\mathbf{L}^\prime)+\lambda_{D-1}(\mathbf{L}^\prime)}\mathbf{L}^\prime$ so that fast consensus among edge servers can be achieved \cite{elsasser2002diffusion}, where $\mathbf{I}$ is the identity matrix, $\mathbf{L}^\prime$ is the product of the Laplacian matrix of graph $\mathcal{G}$ and $\mathbf{\Omega}=\mathrm{diag}(\frac{|\mathcal{S}|}{|\Tilde{\mathcal{S}}_1|},,\dots,\frac{|\mathcal{S}|}{|\Tilde{\mathcal{S}}_D|})$, and $\lambda_i(\mathbf{L}^\prime)$ denotes the $i$-th largest eigenvalue of matrix $\mathbf{L}^\prime$. Once $\hat{\bm{w}}_{k,\alpha}^{\left(d\right)}$ is computed, it updates $\Tilde{\bm{w}}_{k}^{(d)}$ as $\hat{\bm{w}}_{k,\alpha}^{(d)}$, which is then transmitted to the client nodes in $\mathcal{C}_{d}$. The training algorithm is summarized in Algorithm \ref{alg}, where $K$ is assumed to be an integer multiple of $\tau_1\tau_2$. 
\vspace{-3pt}
\begin{algorithm}[t]
\caption{Training Algorithm for SD-FEEL} \label{alg}
\begin{algorithmic}[1]
\STATE{Initialize all client nodes with the same model, i.e., $ \bm{w}_{0}^{\left(i\right)} = \bm{w}_{0},i\in\mathcal{C}$.}
\FOR{$k=1,2,\dots,K$}
	\FOR{each client node $i\in \mathcal{C}$ in parallel}
	\STATE{Update the local model as $\bm{w}_{k}^{\left(i\right)}$ according to \eqref{sgd};}
	\IF{$\mathrm{mod}\left(k,\tau_1\right)=0$}
		\FOR{each edge server $d\in\mathcal{D}$ in parallel}
		\STATE{Collect the most updated model from the client nodes in $\mathcal{C}_{d}$;}
		\STATE{Obtain $\Tilde{\bm{w}}_{k}^{(d)}$ by performing intra-cluster model aggregation according to \eqref{intra};}
	    \IF{$\mathrm{mod}\left(k,\tau_1\tau_2\right)=0$}
	        \STATE{Set $\hat{\bm{w}}_{k,0}^{\left(d\right)}$ as $\Tilde{\bm{w}}_{k}^{(d)}$;}
    		\FOR{$l=1,\dots,\alpha$}
    		\STATE{Send the most updated model $\hat{\bm{w}}_{k,l-1}^{\left(d\right)}$ to its one-hop neighbors $\mathcal{N}_{d}$;}
    		\STATE{Receive models from $\mathcal{N}_{d}$ and perform inter-cluster model aggregation according to \eqref{inter};}
            \ENDFOR 
            \STATE{Update $\Tilde{\bm{w}}^{\left(d\right)}_{k}$ as $\hat{\bm{w}}_{k,\alpha}^{(d)}$;}
		\ENDIF
		\STATE{Broadcast $\Tilde{\bm{w}}^{\left(d\right)}_{k}$ to the client nodes in $\mathcal{C}_{d}$;}
		\ENDFOR
	\ENDIF
	\ENDFOR
\ENDFOR
\end{algorithmic}
\end{algorithm}
\section{Convergence Analysis for SD-FEEL}\label{sec-3}
In this section, we prove the convergence of Algorithm \ref{alg}, and draw various insights from our analysis.
\subsection{Assumptions}
We make the following assumptions on the loss functions to facilitate the convergence analysis.
\begin{assumption}\label{assumptions}
For all $i\in \mathcal{C}$, we assume:
\begin{itemize}[leftmargin=1em]
    \item (\textbf{Smoothness}) The local loss function is $L$-smooth, i.e.,
        \begin{equation}
        \left\| \nabla\! F_i(\!\bm{w}_1\!) \!-\! \nabla\! F_i(\!\bm{w}_2\!) \right\|_2 \!\leq\! L\!\left\| \bm{w}_1\!-\!\bm{w}_2 \right\|_2, \! \forall \bm{w}_1,\!\bm{w}_2\!\in\!\!\mathbb{R}^{\!M}\!.
        \label{eq-smooth}
        \end{equation}
    \item (\textbf{Unbiased and bounded gradient variance}) The mini-batch gradient $g\left(\bm{\xi}; \bm{w}\right)$ is unbiased, i.e.,
        \begin{equation}
            \mathbb{E}_{\bm{\xi}|\bm{w}} [g(\bm{\xi};\bm{w})] = \nabla F_i(\bm{w}), \forall \bm{w}\in\mathbb{R}^M,
        \label{eq-gradient}
        \end{equation}
        and there exists $\sigma>0$ such that
        \begin{equation}
            \mathbb{E}_{\bm{\xi}|\bm{w}} \left[\left\| g(\bm{\xi};\bm{w}) - \nabla F_i(\bm{w}) \right\|_2^2\right] \leq \sigma^2, \forall \bm{w}\in\mathbb{R}^M.
        \label{eq-variance}
        \end{equation}
    \item (\textbf{Degree of non-IIDness}) There exists $\kappa>0$ such that
        \begin{equation}
            \left\| \nabla F_i(\bm{w}) - \nabla F(\bm{w}) \right\|_2 \leq \kappa, \quad \forall \bm{w} \in\mathbb{R}^M.
        \label{kappa}
        \end{equation}
\end{itemize}
\end{assumption}
When $\kappa = 0$, $\nabla F(\bm{w})=\nabla F_i(\bm{w}), \forall i\in\mathcal{C}$, i.e., the assumption in \eqref{kappa} reduces to the case of IID data across the client nodes.
\subsection{Convergence Analysis}
Denote $\mathbf{W}_k\triangleq [\bm{w}_k^{(i)}]\in\mathbb R^{M\!\times\!C}$ and $\mathbf{G}_k\triangleq [g(\bm{\xi}_{k+1}^{(i)}; \bm{w}_k^{(i)})]\in\mathbb R^{M\!\times\!C}$. To characterize the evolution of $\mathbf{W}_k$ as shown in the following lemma, we define $\mathbf{V} \triangleq [v^{i,d}] \in \mathbb{R}^{C\times D}$ and $\mathbf{B} \triangleq [b^{d,i}] \in \mathbb R^{D\times C}$, where $v^{i,d}= \frac{\left|\mathcal{S}_i\right|}{\left| \Tilde{\mathcal{S}}_d \right|} \mathbbm{1}\left\{ i\in \mathcal{C}_d \right\}$, $b^{d,i} = \mathbbm{1}\left\{ i\in \mathcal{C}_d \right\}$, and $\mathbbm{1}\{\cdot\}$ is the indicator function.
\begin{lemma}\label{lemma-1}
    The local models evolve according to the following expression:
    \begin{equation}
    \mathbf{W}_{k+1} = (\mathbf{W}_k -\eta \mathbf{G}_k)\mathbf{T}_k,\, k=1,2,\dots,K,
    \label{eq-relation}
    \end{equation}
    where
    \begin{equation}
        \mathbf{T}_k \! = \! \left\{
        \begin{array}{ll}
        \!\mathbf{V}\mathbf{B},
        & \text{if}\;\mathrm{mod}\left(k,\!\tau_1\right)\!=\!0\; 
        \text{and}\;\mathrm{mod} \left(k,\!\tau_1\tau_2\right)\!\neq\!0, \\
        \!\mathbf{V}\mathbf{P}^\alpha\mathbf{B},
        & \text{if}\; \mathrm{mod}\left(k,\!\tau_1\tau_2\right)=0, \\
        \mathbf{I}, & \text{otherwise.}
        \end{array}
        \right.
        \label{T-k}
    \end{equation}
\end{lemma}
\noindent\emph{Proof.}\;  The proof can be obtained by rewriting \eqref{inter} as $\hat{\mathbf{W}}_{k,l} = \mathbf{P} \hat{\mathbf{W}}_{k,l-1}$ and showing $\hat{\mathbf{W}}_{k,\alpha} = \mathbf{P}^\alpha \hat{\mathbf{W}}_{k,0}$. Details are omitted for brevity.
\QEDB

For convenience, we define $\bm{u}_k\triangleq \sum_{i\in\mathcal{C}} m_i \bm{w}_{k}^{(i)}$ where $m_i\triangleq\frac{|\mathcal{S}_i|}{|\mathcal{S}|}$, and $\left\|\mathbf{X}\right\|_{\mathbf{M}} \triangleq \sum_{i=1}^M\sum_{j=1}^N m_{i,j} |x_{i,j}|^2$ is the weighted Frobenius norm of an $M\times N$ matrix $\mathbf{X}$. Following Lemma 8 in \cite{tang2018d} and leveraging the evolution expression of $\mathbf{W}_k$ in \eqref{eq-relation}, we bound the expected change of the local loss functions in consecutive iterations as follows:
\begin{equation*}
    \mathbb{E}[F\!(\bm{u}_{k+1})] \!-\! \mathbb{E}[F\!(\bm{u}_{k})]
    \!\leq\! -\! \frac{\eta}{2} \mathbb{E} \left[\left\| \nabla \!F(\bm{u}_k) \right\|_2^2 \right]
    \!+\! \frac{\eta^2L}{2} \sum_{i\in\mathcal{C}} m_i^2 \sigma^2
    \setlength{\belowdisplayskip}{-3pt}
\end{equation*}
\begin{equation}
    \!-\!(\frac{\eta}{2}-\!\frac{\eta^2 L}{2}\!) J_k
    \!+\!\frac{\eta L^2}{2}\! \mathbb{E} \!\left[\!\left\| \mathbf{W}_k(\mathbf{I}\!-\!\mathbf{M})\right\|_{\mathbf{M}}^2\!\right]\!,
    \label{one-step}
\end{equation}
where $\mathbf{M}\triangleq \bm{m}\bm{1}^T$ and $J_k \!\triangleq\! \mathbb{E}\! \left[ \left\| \sum_{i \in \mathcal{C}} m_i \nabla F_i(\bm{w}_k^{\!(i)}) \! \right\|_{\!2}^{\!2} \right]$. It remains to bound the last term in the right-hand side (RHS) of \eqref{one-step} in order to show the convergence of Algorithm \ref{alg}, which can be interpreted as the deviation of the local models from their mean, as shown in Lemma \ref{lemma-2}.
\begin{lemma}\label{lemma-2}
With Assumption \ref{assumptions}, we have:
\begin{equation}
\begin{split}
\frac{1}{K} \sum_{k=1}^{K} \mathbb E \left[\!\left\| \mathbf{W}_k(\mathbf{I}-\mathbf{M}) \right\|_{\mathbf{M}}^2 \!\right]
& \leq \frac{8\eta^2 V_2}{K} \sum_{k=1}^{K} J_k \\
& + 2\eta^2 V_1 \sigma^2 + 8\eta^2 V_2 \kappa^2,
\label{eq-lemma2}
\end{split}
\end{equation}
where $\zeta \!=\! |\lambda_2(\mathbf{P})| \!\in\! [0,1)$, $\Lambda \!\triangleq\! \frac{\zeta^{2\alpha}}{1-\zeta^{2\alpha}} \!+\! \frac{2\zeta^\alpha}{1-\zeta^\alpha} \!+\! \frac{\zeta^{2\alpha}}{(1-\zeta^\alpha)^2}$, $V_3 \triangleq \tau_1\tau_2 \left(\tau_1\tau_2\Lambda + \frac{\tau_1\tau_2-1}{2} \frac{2-\zeta^\alpha}{1-\zeta^\alpha} \right)$, $V_1 \triangleq \left(\tau_1\tau_2 \frac{\zeta^{2\alpha}}{1-\zeta^{2\alpha}} + \frac{\tau_1\tau_2-1}{2} \right)$ $/(1-16\eta^2L^2V_3)$, and $V_2 \triangleq V_3/(1-16\eta^2L^2V_3)$.
\end{lemma}
\noindent\emph{Proof.}\; By using \eqref{eq-relation}, we expand $\mathbf{W}_{\!k}(\mathbf{I}-\mathbf{M})$ as $\mathbf{W}_0(\mathbf{I}-\mathbf{M}) \prod_{l=0}^{k-1}\mathbf{T}_{l} - \eta \sum_{s=0}^{k-1}\mathbf{G}_{s}\left( \prod_{l=s}^{k-1}\mathbf{T}_{l} \!-\! \mathbf{M} \right)$. Since $\bm{w}_{0}^{(i)}=\bm{w}_{0}, \forall i\in\mathcal{C}$, \eqref{eq-lemma2} can be shown by bounding $\mathbb E \left[\left\| \eta \sum_{s=0}^{k-1}\mathbf{G}_{s}\left( \prod_{l=s}^{k-1}\mathbf{T}_{l} - \mathbf{M} \right) \right\|_{\mathbf{M}}^2 \right]$. Please refer to Appendix \ref{appendix-lemma-2} for the complete proof.
\QEDB

By substituting \eqref{T-k} in the RHS of \eqref{eq-relation} and choosing a proper learning rate to eliminate the term $\frac{1}{K} \!\sum_{k=1}^{K} J_k$, we are able to bound the expected average-square gradients in the following theorem.
\begin{theorem}
    \label{thm-1}
    If the learning rate $\eta$ satisfies:
    \begin{equation}
        1-\eta L-8\eta^2L^2 V_2\geq 0, 1-16\eta^2L^2V_3 > 0,
        \setlength{\belowdisplayskip}{-3pt}
        \label{lr}
    \end{equation}
    we have:
    \begin{equation}
    \begin{split}
    \frac{1}{K} \sum_{k=1}^{K} \mathbb{E} \left[\left\| \nabla F(\bm{u}_k) \right\|_2^2\right]
    &\leq \frac{2\Delta}{\eta K} + \eta L \sum\nolimits_{i\in\mathcal{C}} m_i^2 \sigma^2 \\
    & + 2\eta^2 L^2 V_1\sigma^2 + 8\eta^2 L^2 V_2 \kappa^2,
    \end{split}
    \label{theorem}
    \end{equation}
    where $\Delta \triangleq \mathbb{E}\left[F(\bm{u}_1)\right]- \mathbb{E}\left[F(\bm{u}^*)\right]$ and $\bm{u}^*\triangleq \arg\min_{\bm{w}}F(\bm{w})$.
\end{theorem}
\noindent\emph{Proof.}\; Please refer to the Appendix \ref{appendix-thm}.\QEDB

\begin{figure}[t]
    \centering
    \includegraphics[width=0.9\linewidth]{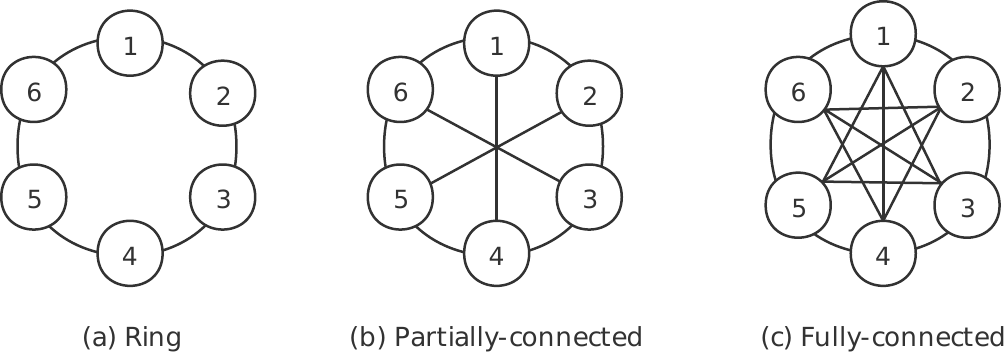}
    \caption{Typical network topologies of the edge servers.}
    \label{Topology}
\end{figure}

With the results in Theorem \ref{thm-1}, we are able to draw various insights as elaborated in the following remarks.
\begin{remark}\label{remark-1}
It is straightforward that the RHS of \eqref{theorem} increases with both $\tau_1$ and $\tau_2$. Thus, it is preferable to choose $\tau_1=\tau_2=1$ to reduce the number of training iterations without considering the training time. However, as will be seen in Section \ref{sec-4}, performing either intra-cluster or inter-cluster model aggregations in the most frequent way turns out to be sub-optimal for the overall training latency.
\end{remark}
\begin{remark}\label{remark-2}
By taking the first-order derivative, we see that the RHS of \eqref{theorem} also increases with $\zeta^\alpha$, indicating that either increasing $\alpha$ (i.e., increasing the inter-server communication overhead) or decreasing $\zeta$ (i.e., increasing the degree of connectivity among the edge servers) results in faster convergence. For better illustration, three typical network topologies of six edge servers are shown in Fig. \ref{Topology}, including the ring, partially connected and fully connected topologies, and the values of $\zeta$ are given as 0.6, 0.333 and 0, respectively. In other words, the fully-connected network topology achieves the best performance. Besides, in the extreme case with $\alpha \!\rightarrow\! \infty$, the edge servers can reach consensus after inter-cluster model aggregation (i.e., $\Tilde{\bm{w}}_{k}^{(d)} \!\leftarrow\! \sum_{j\in\mathcal{D}} \frac{|\Tilde{\mathcal{S}}_j|}{|\mathcal{S}|} \Tilde{\bm{w}}_{k}^{(j)}$) even with a sparsely connected network.
\end{remark}

\begin{remark}\label{remark-3}
When $\tau_1=1$, $\tau_2=1$, and $\zeta^\alpha=0$, i.e., the local models of all the client nodes are synchronized at each iteration, the convergence result in Theorem \ref{thm-1} reduces to that of the fully synchronous SGD algorithm \cite{parallelsgd}. In addition, when $\kappa=0$, i.e., in the IID case, our result also improves over that obtained in \cite{castiglia2020multi}.
\end{remark}
\section{Experiments}\label{sec-4}
\subsection{Settings}
We simulate an SD-FEEL system with 50 client nodes and 10 edge servers, and each edge server is associated with five client nodes. In our experiments, we consider a ring topology of the edge servers as shown in Fig. \ref{Topology}(a) unless otherwise specified.

SD-FEEL is evaluated on the CIFAR-10 dataset, which is a benchmark dataset for image classification tasks containing 10 categories of images. To simulate the non-IID data distribution, we follow \cite{wang2020federated} by allocating a proportion $p_{l,j}$ of the training samples of class $l$ to the $j$-th client node, where $p_{l,j}$ is sampled from a Dirichlet distribution $\text{Dir}_{50}(0.5)$. We use a convolutional neural network  with two $5\times 5$ convolution layers that consists of 21,840 trainable parameters \cite{liu2020client}. The batch size and the learning rate of mini-batch SGD are set as 10 and 0.01, respectively.

To verify the effectiveness of SD-FEEL in reducing training latency, we adopt the following FL schemes for comparisons:
\begin{itemize}[leftmargin=1em]
    \item FedAvg \cite{mcmahan2017communication}: This is a Cloud-based FL scheme, where a Cloud-based PS collects and aggregates models from all client nodes after every $\tau\!=\!\tau_1\tau_2$ local model updates.
    \item FEEL \cite{lim2020federated}: This is a conventional edge-assisted FL scheme, where an edge server randomly schedules five client nodes at each iteration due to the limited number of wireless channels. Note that this is consistent with the number of accessible client nodes for each edge server in SD-FEEL.
    \item HierFAVG \cite{liu2020client}: This FL scheme leverages a Cloud-based PS and multiple edge servers. Specifically, the edge servers perform local model aggregation after every $\tau_1$ local updates; whereas the Cloud-based PS aggregates models from the edge servers after every $\tau_2$ local model aggregations.
\end{itemize}
\subsection{Training Latency Calculation}
For SD-FEEL, the latency of the $K$ training iterations is calculated as $t_\text{tot} \!=\! \left\lceil\frac{K}{\tau_1\tau_2}\right\rceil \left[ \tau_2 ( \tau_1 t_\text{cmp}^\text{ct} + t_\text{cmm}^\text{ct-sr} ) + \alpha t_\text{cmm}^\text{sr-sr} \right]$, where $t_\text{cmp}^\text{ct}$ denotes the computation latency for each local update, $t_\text{cmm}^\text{ct-sr}$ denotes the model uploading latency between a client node and its associated edge server, and $t_\text{cmm}^\text{sr-sr}$ is the model transmission latency between neighboring edge servers. We assume $t_\text{cmp}^\text{ct} \!=\! \frac{N_\text{MAC}}{C_\text{CPU}}$, where $N_\text{MAC} \!=\! 138.4\,\text{MFLOPs}$ is the number of the floating-point operations (FLOPs) required for each local iteration\footnote{The number of FLOPs required for local training at each iteration is calculated using OpCounter, which is an open-source model analysis library available at https://github.com/Lyken17/pytorch-OpCounter.}, and $C_\text{CPU} \!=\! 10 \,\text{GFLOPS}$ denotes the computing bandwidth of the local processors. The communication latency is expressed as $t_\text{cmm} \!=\! \frac{M_{\rm{bit}}}{R}$ with $M_{\rm{bit}} \!=\! 32M \,\text{bits}$ and $R$ denoting the transmission rate. Specifically, we assume that the client nodes communicate with the associated edge servers using orthogonal channels and there is no inter-cluster interference \cite{shi2020joint}. The transmission rate is assumed to be $R^\text{ct-sr}\!=\!B\log_2{(1+\text{SNR}}) \!\approx\! 5\,\text{Mbps}$, where $B\!=\!1\,\text{MHz}$ and $\text{SNR}\!=\!15\,\text{dB}$. The edge server communicates with the neighboring edge servers via high-speed links with the bandwidth of $50\,\text{Mbps}$ \cite{hu2020coedge}, and the bandwidth from the edge servers to the Cloud is set as $5\,\text{Mbps}$. Accordingly, the transmission rate from the client nodes to the Cloud is given by $R^\text{ct-cd}\!=\!2.5\,\text{Mbps}$.

\subsection{Results}
\subsubsection{Convergence Speed and Learning Performance}
\begin{figure}[t]
\setlength\abovecaptionskip{-5pt}
\setlength\belowcaptionskip{-8pt}
    \centering
    \subfigure
    {\includegraphics[width=0.48\linewidth]{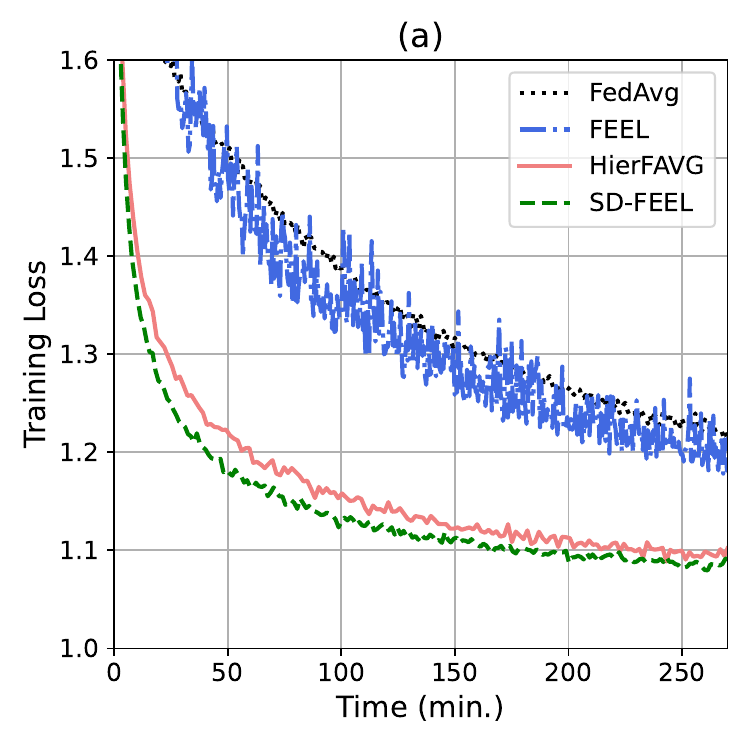}}
    \subfigure
    {\includegraphics[width=0.48\linewidth]{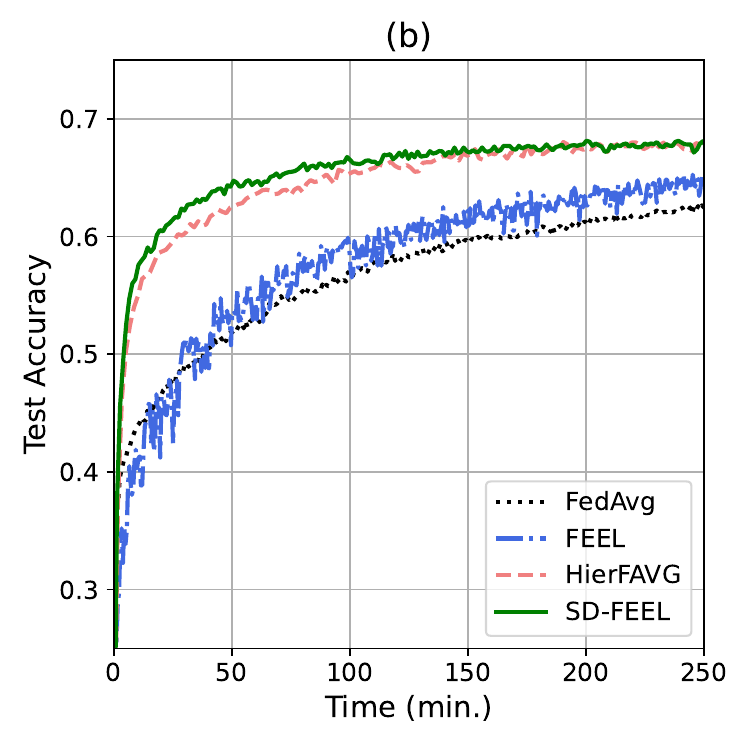}}
    \caption{(a) Training loss and (b) test accuracy over time ($\tau_1=2$, $\tau_2=1$ and $\alpha=5$).}
    \label{fig:convergence}
\end{figure}

We show the training loss and test accuracy over time in Fig. \ref{fig:convergence}(a) and Fig. \ref{fig:convergence}(b), respectively. It can be observed that, the training loss of SD-FEEL drops rapidly at the early stage of the training process, and converges at around 250 minutes. Comparatively, the training progress of FedAvg lags far behind due to slow model uploading from the client nodes to the Cloud-based PS in each iteration. On the other hand, due to infrequent communications to the Cloud, HierFAVG also converges within 250 minutes. However, it suffers higher training loss compared to SD-FEEL since the communication among edge servers in SD-FEEL is more efficient and thus accelerates the learning progress. It is also worthwhile to note that, although FEEL enjoys a low uplink transmission delay, the limited number of accessible client nodes results in an unstable training process and thus leads to much higher training loss. This demonstrates the effectiveness of SD-FEEL in achieving fast convergence by removing the reliance on the Cloud-based infrastructure. From Fig. \ref{fig:convergence}(b), it is seen that SD-FEEL secures test accuracy improvement compared to HierFAVG, while the learned models of FedAvg and FEEL within the given training time are simply unusable. 
\subsubsection{Ablation Study}
\begin{figure}[t]
\setlength\abovecaptionskip{-5pt}
\setlength\belowcaptionskip{-8pt}
    \centering
    \subfigure
    {\includegraphics[width=0.48\linewidth]{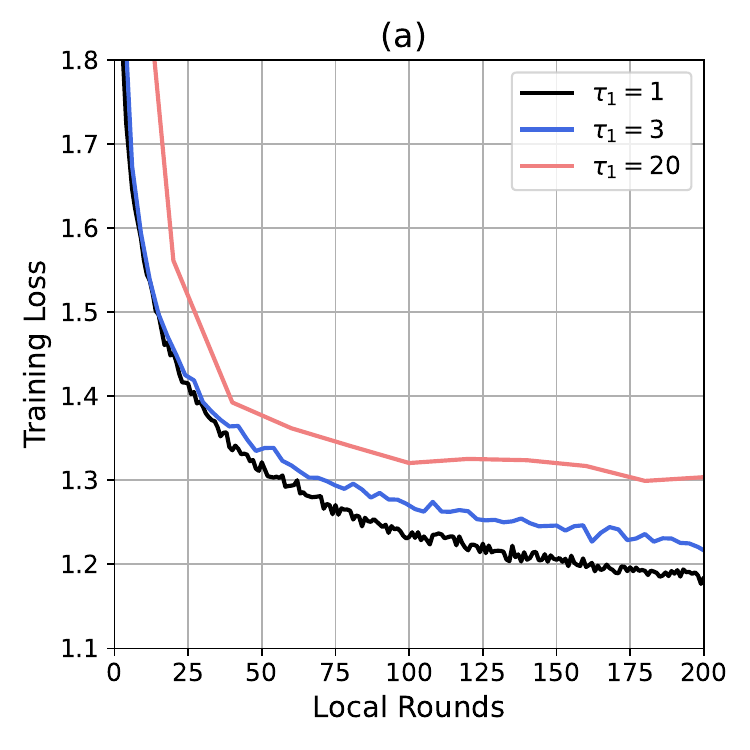}}
    \subfigure
    {\includegraphics[width=0.48\linewidth]{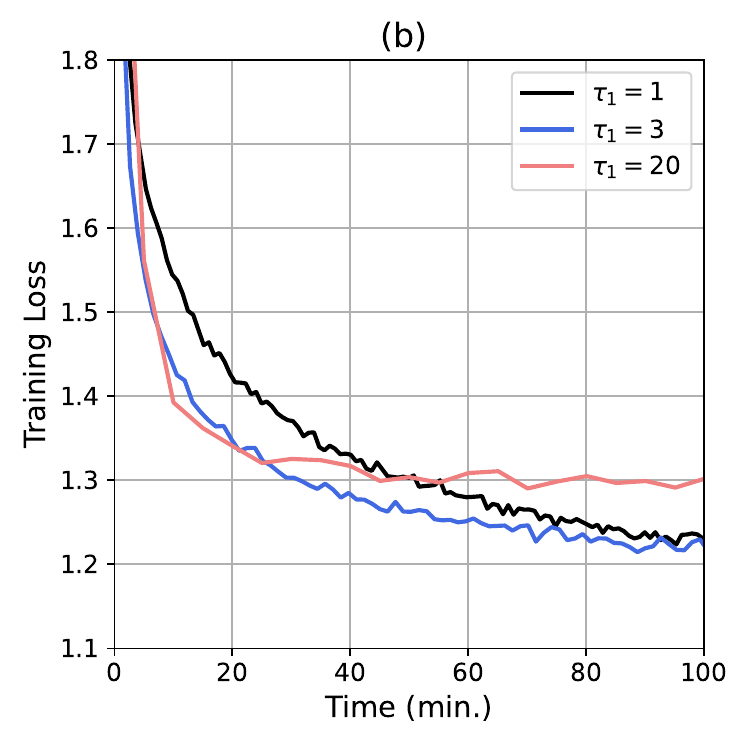}}
    \caption{Training loss of SD-FEEL ($\tau_2=1$ and $\alpha=1$) over (a) iterations and (b) time.}
    \label{fig:k1}
\end{figure}

We investigate the impacts of the intra-cluster model aggregation period $\tau_1$ by showing the relationship between training loss and iterations (training latency) in Fig. \ref{fig:k1}(a) (Fig. \ref{fig:k1}(b)). For both figures, we fix $\tau_2$ to 1 and evaluate the configurations with $\tau_1\!=\!1, 3,\,\text{and}\,20$. From Fig. \ref{fig:k1}(a) we observe that a smaller value of $\tau_1$ leads to a lower training loss within a given number of training iterations, as discussed in Remark \ref{remark-1}. However, this conclusion becomes invalid in terms of training latency as shown in Fig. \ref{fig:k1}(b), where $\tau_1\!=\!3$ achieves the minimum training loss. This is owing to the fact that less frequent intra-cluster model aggregations help to save communication time. Similar behaviors can be observed by varying $\tau_2$ with a given value of $\tau_1$. We omit the results here due to the limited space. These observations jointly necessitate an optimal choice of the intra-/inter-cluster model aggregation frequencies for the minimum training latency.

\begin{figure}[t]
\setlength\abovecaptionskip{-5pt}
\setlength\belowcaptionskip{-8pt}
    \centering
    \subfigure
    {\includegraphics[width=0.48\linewidth]{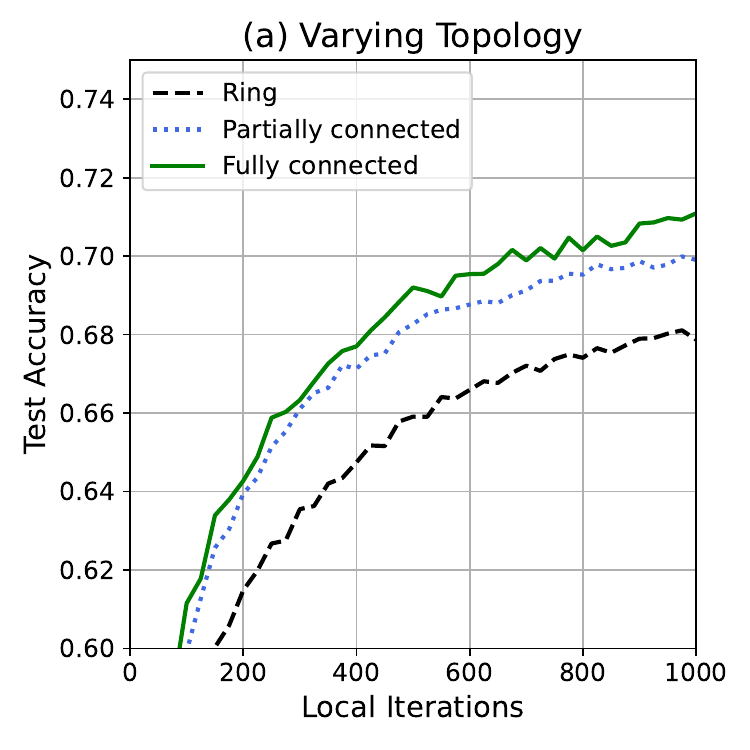}}
    \subfigure
    {\includegraphics[width=0.48\linewidth]{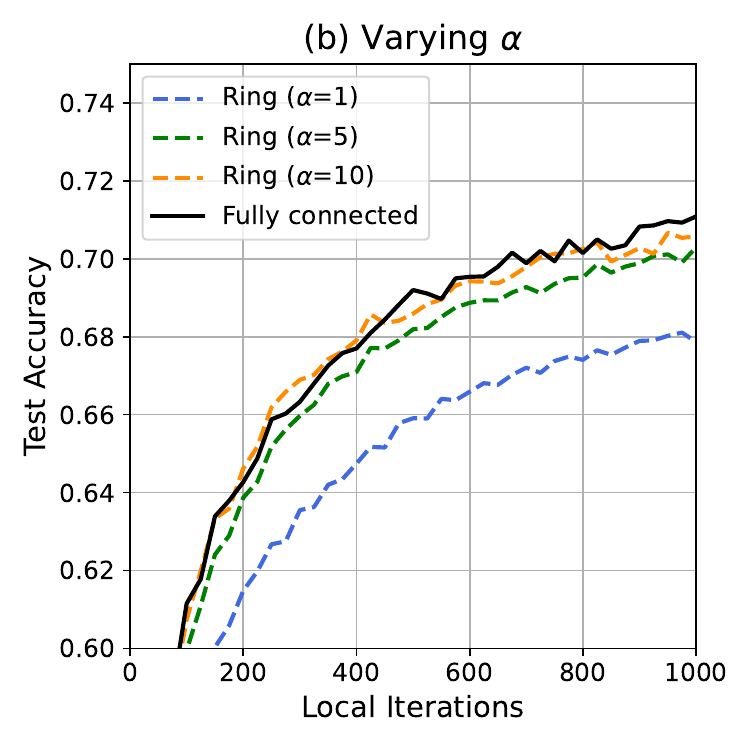}}
    \caption{Test accuracy over iterations ($\tau_1=5$, $\tau_2=5$, and $\alpha=1$ by default) with (a) different network topologies and (b) different values of $\alpha$.}
    \label{fig:topo}
\end{figure}

We also evaluate the test accuracy of SD-FEEL on different network topologies of the edge servers. As shown in Fig. \ref{fig:topo}(a), we see that within a given number of training iterations, a more connected network topology achieves a higher test accuracy, since more information is collected from neighboring edge clusters in each round of inter-server model aggregation, as explained in Remark \ref{remark-2}. Besides, effects of $\alpha$ on the test accuracy are examined via a ring topology in Fig. \ref{fig:topo}(b). In line with Remark \ref{remark-2}, a larger value of $\alpha$ increases the training speed in terms of iterations. It is also observed that when $\alpha$ is greater than 10, the test accuracy approaches the case with a fully-connected network, and further increasing $\alpha$ brings marginal improvement. Therefore, based on the network topology of edge servers, we can choose a suitable value of $\alpha$ to balance the communication cost and learning performance.

\begin{figure}[t]
\setlength\abovecaptionskip{-5pt}
\setlength\belowcaptionskip{-8pt}
    \centering
    \subfigure
    {\includegraphics[width=0.48\linewidth]{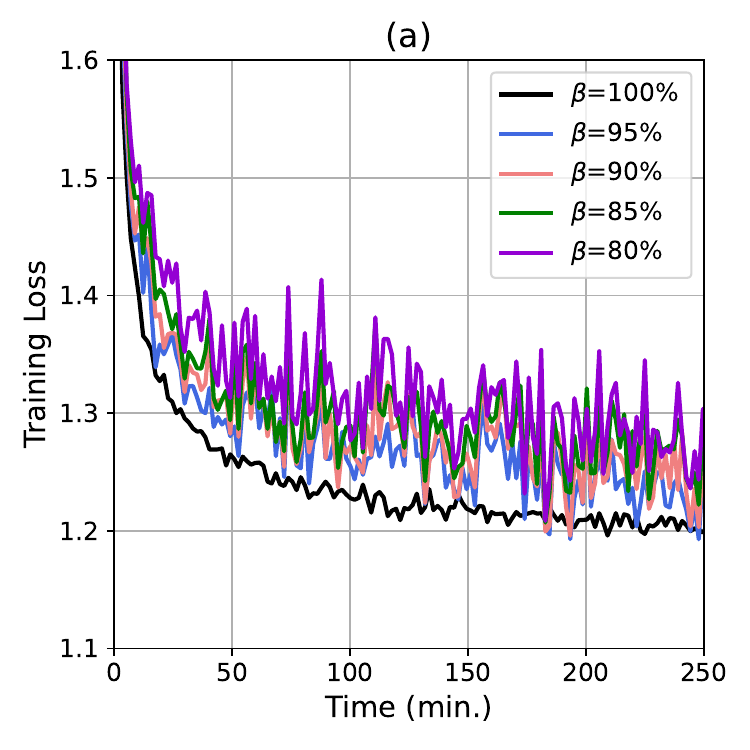}}
    \subfigure
    {\includegraphics[width=0.48\linewidth]{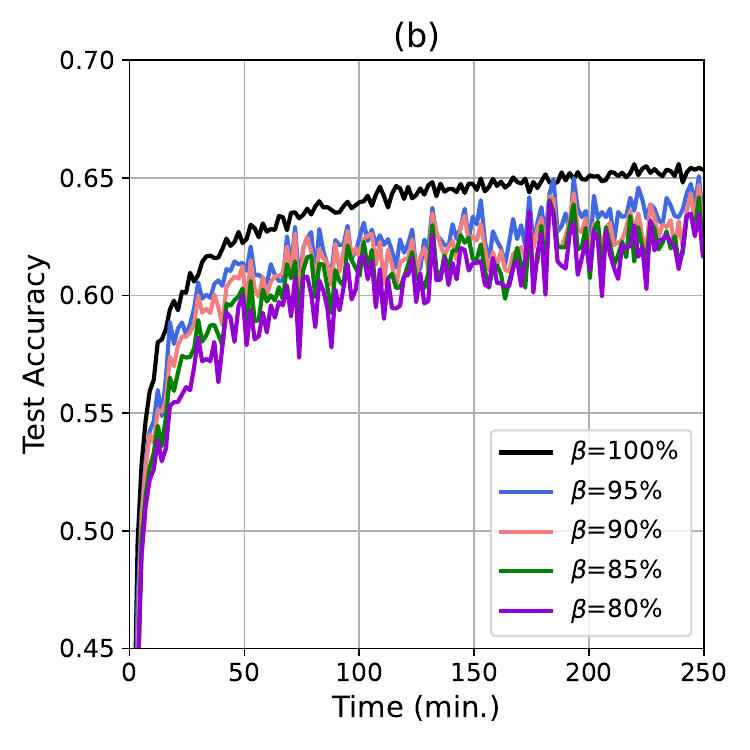}}
    \caption{(a) Training loss and (b) test accuracy over time ($\tau_1=5$, $\tau_2=1$, and $\alpha=1$) with different values of $\beta$.}
    \label{fig:partial}
\end{figure}

We further evaluate SD-FEEL with probabilistic dropout of the client nodes in Fig. \ref{fig:partial}, where a client node participates the training process with a certain probability, denoted as $\beta$, in each round. It is observed that SD-FEEL still maintains the convergence behavior with a reasonable amount of dropout client nodes, despite experiences slight degradation of the learned model accuracy. We leave the thorough investigation on SD-FEEL with partial participation to our further works.
\section{Conclusions}\label{sec-5}
In this paper, we investigated a novel FL architecture, namely semi-decentralized federated edge learning (SD-FEEL), to realize low-latency distributed learning on non-IID data. Convergence analysis was conducted for the training algorithm of SD-FEEL, from which, various insights were drawn for system implementation. Simulations demonstrated that SD-FEEL substantially reduces the training latency while improving the learning performance. For future works, it is worth investigating the optimal intra-cluster and inter-cluster model aggregation frequencies and analyzing the convergence of SD-FEEL with partial participation. It would also be interesting to consider more practical scenarios with device heterogeneity. 


\newpage
\onecolumn
\appendix
\section{Proof of Main Result}
\subsection{Notations}
We denote $m_{i} \!\triangleq\! \frac{|\mathcal{S}_i|}{|\mathcal{S}|}$, $\hat{m}_i \triangleq \frac{|\mathcal{S}_i|}{|\Tilde{\mathcal{S}}_d|}$, $\bm{m} \!\triangleq\! [m_{i}] \!\in\! \mathbb R^{C}$, $\mathbf{M} \!\triangleq\! \bm{m}\bm{1}^T$, $\mathbf{H}_1 \!\triangleq\! \mathbf{V}\mathbf{B}$, $\mathbf{H}_2 \!\triangleq\! \mathbf{V} \mathbf{P}^\alpha \mathbf{B}$, and $\nabla \mathbf{\Tilde{F}}_k \!\triangleq\! \left[\nabla F_1 (\bm{w}_k^{(1)}), \dots, \nabla F_C (\bm{w}_k^{(C)})\right]$. 
We also define the accumulated stochastic gradient and derivative of the global loss function as follows: 
\begin{align*}
Y_{j,l}^{(1)} &\triangleq \sum_{s=j\tau_2\tau_1+1}^{j\tau_2\tau_1+l\tau_1} \mathbf{G}_s, &
Y_{j,l,f}^{(2)} &\triangleq \sum_{s=j\tau_2\tau_1+l\tau_1+1}^{j\tau_2\tau_1+l\tau_1+f} \mathbf{G}_s, &
Y_r &\triangleq \sum_{s=r\tau_2\tau_1+1}^{(r+1)\tau_2\tau_1} \mathbf{G}_s, \\
Q_{j,l}^{(1)} &\triangleq \sum_{s=j\tau_2\tau_1+1}^{j\tau_2\tau_1+l\tau_1} \nabla \mathbf{\Tilde{F}}_s, &
Q_{j,l,f}^{(2)} &\triangleq \sum_{s=j\tau_2\tau_1+l\tau_1+1}^{j\tau_2\tau_1+l\tau_1+f} \nabla \mathbf{\Tilde{F}}_s, &
Q_{r} &\triangleq \sum_{s=r\tau_2\tau_1+1}^{(r+1)\tau_2\tau_1} \nabla \mathbf{\Tilde{F}}_s.
\end{align*}
\subsection{Lemmas}
\begin{lemma}\label{proposition}
The eigenvalues of $\mathbf{H}_1$ and $\mathbf{H}_2$ have the following properties:
\begin{itemize}
    \item $\mathbf{H}_1$ has a left eigenvector $\bm{1}^T$ with eigenvalue 1, and a right eigenvector $\bm{m}$ with eigenvalue 1.\\
    \item $\mathbf{H}_2$ has a left eigenvector $\bm{1}^T$ with eigenvalue 1, and a right eigenvector $\bm{m}$ with eigenvalue 1. \\
    \item The eigenvalues of $\mathbf{H}_2$ are same as the eigenvalues of $\mathbf{P}^\alpha$, and the absolute values of all the eigenvalues except the largest one are no greater than 1.
\end{itemize}
\end{lemma}
\noindent\emph{Proof.}\; Since when $\alpha=0$, $\mathbf{H}_2$ reduces to $\mathbf{H}_1$, it suffices to prove the first and second properties for $\mathbf{H}_2$ for arbitrary $\alpha$. 
Denote the associated edge server of client node $i$ as $d(i)$.
According to the definition of $\mathbf{H}_2 \triangleq [ h_{2}^{i,j} ]$, we have:
\begin{equation}
    h_{2}^{i,j} = \sum_{d^\prime}(\sum_{d} v^{i,d} p_{\alpha}^{d,d^\prime}) b^{d^\prime,j}
    \overset{\text{(a)}}{=} v^{i,d(i)}  p_{\alpha}^{d(i),d(j)} b^{d(j),j}
    = \hat{m}_i  p_{\alpha}^{d(i),d(j)},
\label{eq-h2}
\end{equation}
where $p_{\alpha}^{d,d^\prime}$ denotes the element in $d$-th row and $d^\prime$-th column of $\mathbf{P}^\alpha$, (a) holds since $v^{i,d}= \hat{m}_i \mathbbm{1}\left\{ i\in \mathcal{C}_d \right\}$ and $b^{d,i} = \mathbbm{1}\left\{ i\in\mathcal{C}_d \right\}$. Since $\sum_{i} h_{2}^{i,j}=\sum_{d\in\mathcal{D}} (\sum_{i\in\mathcal{C}_d} \hat{m}_i p_{\alpha}^{d(i),d(j)})= \sum_{d\in\mathcal{D}} p_{\alpha}^{d,d^\prime}=1$, it is easy to verify that $\mathbf{H}_2$ has a left eigenvector $\bm{1}^T$ with eigenvalue 1. Similarly, since $\sum_{j} h_{2}^{i,j} m_j = \hat{m}_i (\sum_{j} p_{\alpha}^{d(i),d(j)} m_j)=\hat{m}_i$, $\mathbf{H}_2$ has a right eigenvector $\bm{m}$ with eigenvalue 1.

By invoking Proposition 2 of \cite{castiglia2020multi}, it can be shown that the non-zero eigenvalues of $\mathbf{H}_2$ are same as those of $\mathbf{P}^\alpha$. Since $\mathbf{P}$ is doubly stochastic, i.e., it satisfies $\mathbf{P}^T=\mathbf{P}$ and $\mathbf{P}\bm{1}=\bm{1}$, the $D$ eigenvalues satisfy $1 = |\lambda_1(\mathbf{P})| > |\lambda_2(\mathbf{P})| \geq \dots \geq |\lambda_D(\mathbf{P})| \geq 0$. Because $\lambda_{i}\left(\mathbf{P}^{\alpha}\right)=\lambda_{i}^{\alpha}\left(\mathbf{P}\right), i=1,\cdots, D$, we have $1 = \left| \lambda_{1}\left(\mathbf{P}^{\alpha}\right) \right| > \left| \lambda_{2}\left(\mathbf{P}^{\alpha}\right) \right| \geq \dots \geq \left| \lambda_{D}\left(\mathbf{P}^{\alpha}\right) \right| \geq 0$.
\QEDB
\begin{lemma}\label{lemma-3} 
Matrix $\mathbf{T}_k$ in \eqref{T-k} satisfies $\mathbf{T}_k\mathbf{M}=\mathbf{M}\mathbf{T}_k=\mathbf{M}$.
\end{lemma}
\noindent\emph{Proof.}\; We show this lemma by analyzing the three cases of \eqref{T-k} respectively. For the first case, we have $\mathbf{T}_k\mathbf{M}=\mathbf{V}\mathbf{B}\mathbf{M}=\mathbf{H}_1\bm{m}\bm{1}^T=\bm{m}\bm{1}^T=\mathbf{M}$, and $\mathbf{M}\mathbf{T}_k=\mathbf{M}\mathbf{V}\mathbf{B}=\bm{m}\bm{1}^T\mathbf{H}_1=\bm{m}\bm{1}^T=\mathbf{M}$. For the second case, we have $\mathbf{T}_k\mathbf{M}=\mathbf{V} \mathbf{P}^\alpha \mathbf{B}\mathbf{M}=\mathbf{H}_2\bm{m}\bm{1}^T=\bm{m}\bm{1}^T=\mathbf{M}$, and $\mathbf{M}\mathbf{T}_k=\mathbf{M}\mathbf{V} \mathbf{P}^\alpha \mathbf{B}=\bm{m}\bm{1}^T\mathbf{H}_2=\bm{m}\bm{1}^T=\mathbf{M}$. Moreover, the third case can be shown by similar arguments, i.e. $\mathbf{T}_k\mathbf{M}=\mathbf{I}\mathbf{M}=\mathbf{M}$ and $\mathbf{M}\mathbf{T}_k=\mathbf{M}\mathbf{I}=\mathbf{M}$.
\QEDB
\begin{lemma}\label{lemma-one-step}
The expected change of the global loss functions in consecutive iterations can be bounded as follows:
\begin{equation}
\begin{split}
    \mathbb{E}[F(\bm{u}_{k+1})] - \mathbb{E}[F(\bm{u}_{k})]
    \leq -\frac{\eta}{2} (1 - \eta L) J_k 
    + \frac{\eta L^2}{2} \mathbb{E} \left[\left\| \mathbf{W}_k (\mathbf{I}-\mathbf{M})\right\|_\mathbf{M}^2\right] + \frac{\eta^2L}{2} \sum_{i\in\mathcal{C}} m_i^2  \sigma^2.
    \label{eq-one-step}
\end{split}
\end{equation} 
\end{lemma}
\noindent\emph{Proof.}\; First, we right multiply both sides of the evolution expression in \eqref{eq-relation} by $\bm{m}$, yielding the following expression:
\begin{equation}
    \bm{u}_{k+1} = \bm{u}_k - \eta \mathbf{G}_k\bm{m}.
    \label{eq-u-k}
\end{equation}

We apply the global loss function to both sides of \eqref{eq-u-k}, and due to the $L$-smoothness, we have:
\begin{align}
    \mathbb{E}[F(\bm{u}_{k+1})]
    &\leq \mathbb{E} [F(\bm{u}_{k})] + \mathbb{E} \left\langle \nabla F(\bm{u}_k),-\eta\mathbf{G}_k\bm{m} \right\rangle + \frac{L}{2} \mathbb{E}\left\| \eta \mathbf{G}_k\bm{m} \right\|_2^2 \nonumber \\
    &= \mathbb{E} [F(\bm{u}_{k})] -\eta \mathbb{E}\left\langle \nabla F(\bm{u}_k),\mathbb{E} [\mathbf{G}_k\bm{m}] \right\rangle + \frac{\eta^2L}{2} \mathbb{E}\left\| \mathbf{G}_k\bm{m} - \nabla \mathbf{\Tilde{F}}_k\bm{m} + \nabla \mathbf{\Tilde{F}}_k\bm{m} \right\|_2^2 \nonumber \\
    &\overset{(a)}{=} \mathbb{E} [F(\bm{u}_{k})] - \eta \mathbb{E}\left\langle \nabla F(\bm{u}_k), \nabla \mathbf{\Tilde{F}}_k\bm{m} \right\rangle + \frac{\eta^2L}{2} \mathbb{E}\left\| \mathbf{G}_k\bm{m} - \nabla \mathbf{\Tilde{F}}_k\bm{m} \right\|_2^2 + \frac{\eta^2L}{2} \mathbb{E}\left\| \nabla \mathbf{\Tilde{F}}_k\bm{m} \right\|_2^2 \nonumber \\
    &= \mathbb{E} [F(\bm{u}_{k})] - \eta \mathbb{E}\left\langle \nabla F(\bm{u}_k),\sum_{i\in\mathcal{C}} m_i \nabla F_i (\bm{w}_k^{(i)}) \right\rangle \nonumber \\
    &\qquad + \frac{\eta^2L}{2} \mathbb{E}\left\| \sum_{i\in\mathcal{C}} m_i \left( g(\bm{w}_k^{(i)}) - \nabla F_i(\bm{w}_k^{(i)}) \right) \right\|_2^2 + \frac{\eta^2L}{2} \mathbb{E}\left\| \nabla \mathbf{\Tilde{F}}_k\bm{m} \right\|_2^2 \nonumber \\
    &\overset{(b)}{=} \mathbb{E} [F(\bm{u}_{k})] - \eta \mathbb{E}\left\langle \nabla F(\bm{u}_k), \sum_{i\in\mathcal{C}} m_i \nabla F_i (\bm{w}_k^{(i)}) \right\rangle \nonumber \\
    &\qquad + \frac{\eta^2L}{2} \sum_{i\in\mathcal{C}} m_i^2 \mathbb{E}\left\| g(\bm{w}_k^{(i)}) - \nabla F_i(\bm{w}_k^{(i)}) \right\|_2^2 + \frac{\eta^2L}{2} \mathbb{E}\left\| \nabla \mathbf{\Tilde{F}}_k\bm{m} \right\|_2^2 \nonumber \\
    &\overset{(c)}{\leq} \mathbb{E} [F(\bm{u}_{k})] -\left( \frac{\eta}{2}\mathbb{E}\left\| \nabla F(\bm{u}_k) \right\|_2^2 + \frac{\eta}{2}\mathbb{E} \left\| \sum_{i\in\mathcal{C}} m_i \nabla F_i(\bm{w}_k^{(i)})\right\|_2^2 - \frac{\eta}{2} \mathbb{E} \left\| \nabla F(\bm{u}_k)- \sum_{i\in\mathcal{C}} m_i \nabla F_i(\bm{w}_k^{(i)}) \right\|_2^2 \right) \nonumber \\
    &\qquad+ \frac{\eta^2L}{2} \sum_{i\in\mathcal{C}} m_i^2 \sigma^2 + \frac{\eta^2L}{2} \mathbb{E}\left\| \nabla \mathbf{\Tilde{F}}_k\bm{m} \right\|_2^2 \nonumber \\
    &\overset{(d)}{\leq} \mathbb{E} [F(\bm{u}_{k})] -\left( \frac{\eta}{2}\mathbb{E}\left\| \nabla F(\bm{u}_k) \right\|_2^2 + \frac{\eta}{2}\sum_{i\in\mathcal{C}} m_i \mathbb{E} \left\| \nabla F_i(\bm{w}_k^{(i)})\right\|_2^2 - \frac{\eta}{2} \mathbb{E} \left\| \sum_{i\in\mathcal{C}} m_i \left(\nabla F_i(\bm{u}_k)-\nabla F_i(\bm{w}_k^{(i)}) \right) \right\|_2^2 \right) \nonumber \\
    &\qquad+ \frac{\eta^2L}{2} \sum_{i\in\mathcal{C}} m_i^2  \sigma^2 + \frac{\eta^2L}{2} \mathbb{E} \left\| \sum_{i\in\mathcal{C}} m_i \nabla F_i(\bm{w}_k^{(i)})\right\|_2^2 \nonumber \\
    &\overset{(e)}{=} \mathbb{E} [F(\bm{u}_{k})] -\frac{\eta}{2}\mathbb{E}\left\| \nabla F_i(\bm{u}_k) \right\|_2^2 - \left(\frac{\eta}{2} - \frac{\eta^2L}{2} \right)J_k + \frac{\eta}{2} \sum_{i\in\mathcal{C}} m_i \mathbb{E} \left\| \nabla F_i(\bm{u}_k)-\nabla F_i(\bm{w}_k^{(i)}) \right\|_2^2 + \frac{\eta^2L}{2} \sum_{i\in\mathcal{C}} m_i^2  \sigma^2 \nonumber \\
    &\overset{(f)}{\leq} \mathbb{E} [F(\bm{u}_{k})] -\frac{\eta}{2}\mathbb{E}\left\| \nabla F(\bm{u}_k) \right\|_2^2 - \left(\frac{\eta}{2} - \frac{\eta^2L}{2} \right) J_k + \frac{\eta L^2}{2} \sum_{i\in\mathcal{C}} m_i \mathbb{E} \left\| \bm{u}_k - \bm{w}_k^{(i)} \right\|_2^2 + \frac{\eta^2L}{2} \sum_{i\in\mathcal{C}} m_i^2  \sigma^2 \nonumber \\
    &= \mathbb{E} [F(\bm{u}_{k})] -\frac{\eta}{2} (1 - \eta L) J_k 
    + \frac{\eta L^2}{2} \mathbb{E} \left\| \mathbf{W}_k (\mathbf{I}-\mathbf{M})\right\|_\mathbf{M}^2 + \frac{\eta^2L}{2} \sum_{i\in\mathcal{C}} m_i^2 \sigma^2,
    \label{eq-f-u-k}
\end{align}
where (a) follows $\mathbb{E} [\mathbf{G}_k\bm{m}] = \nabla \mathbf{\Tilde{F}}_k\bm{m}$. Since $\mathbb{E}\left[g(\bm{w}_k^{(i)})\right]=\nabla F_i(\bm{w}_k^{(i)})$, the cross-terms of $\mathbb{E}\left\langle m_i g(\bm{w}_k^{(i)}) - m_i\nabla F_i(\bm{w}_k^{(i)}),\right.$ $\left.m_j g(\bm{w}_k^{(j)}) - m_j\nabla F_j(\bm{w}_k^{(j)}) \right\rangle$ are zero and thus (b) holds. (c) follows the assumption in \eqref{eq-variance} and $\bm{a}^T\bm{b}=\frac{1}{2}\left\|\bm{a}\right\|_2^2+\frac{1}{2}\left\|\bm{b}\right\|_2^2-\frac{1}{2}\left\|\bm{a}-\bm{b}\right\|_2^2$, and (d) is due to Jensen's inequality. (e) is according to the definition of $J_k$, and (f) holds because of the $L$-smoothness assumption of the local loss function in \eqref{eq-smooth}. Finally, the proof is completed by moving $\mathbb{E} [F(\bm{u}_{k})]$ to the LHS of \eqref{eq-f-u-k}.
\QEDB
\begin{lemma}\label{lemma-H}  
$\mathbf{H}_1$ and $\mathbf{H}_2$ have the following properties:
\begin{align}
    & \left\|\mathbf{H}_1-\mathbf{M} \right\|_{\mathrm{op}}=1, \\
    & \left\|\mathbf{H}_2^j-\mathbf{M} \right\|_{\mathrm{op}}=\zeta^{j\alpha},
\end{align}
where $\left\|\cdot\right\|_{\mathrm{op}} \triangleq \max_{\left\|\bm{w}\right\|=1} \mathbf{X}\bm{w} = \sqrt{\lambda_\mathrm{max}(\mathbf{X}^T\mathbf{X})}$ denotes the operator norm.
\end{lemma}
\noindent\emph{Proof.}\; According to Lemma \ref{proposition} and the definition of the matrix operation norm, we have:
\begin{equation}
\begin{split}
    \left\|\mathbf{H}_2^j-\mathbf{M} \right\|_{\text{op}} 
    = \sqrt{\lambda_{\max}((\mathbf{H}_2^j-\mathbf{M})^T(\mathbf{H}_2^j-\mathbf{M}))} 
    = \sqrt{\lambda_{\max}(\mathbf{H}_2^{2j}-\mathbf{M}^2)}
    = \sqrt{\lambda_{\max}(\mathbf{H}_1-\mathbf{M})^{2j}}
    = \zeta^{j\alpha}.
    \label{eq-30}
\end{split}
\end{equation}
In particular, by plugging $\alpha=0$ in \eqref{eq-30}, we have $\left\|\mathbf{H}_1-\mathbf{M} \right\|_{\text{op}}=\zeta^{j0}=1$.
\QEDB
\begin{lemma}\label{lemma-global-gradient} 
The gradient of the global loss function can be bounded as follows:
\begin{equation}
    \mathbb{E} \left\| \nabla \mathbf{\Tilde{F}}(\mathbf{W}_k) \right\|_\mathbf{M}^2 \leq 8L^2 E_k + 4\kappa^2 + 4J_k,
    \label{eq-lemma-global-gradient}
\end{equation}
where $E_k \triangleq \mathbb E \left[\left\| \mathbf{W}_k(\mathbf{I}-\mathbf{M}) \right\|_\mathbf{M}^2 \right]$ and $J_k \triangleq \mathbb{E} \left[\left\| \sum_{i\in\mathcal{C}} m_i \nabla F_i(\bm{w}_k^{(i)})\right\|_2^2 \right]$.
\end{lemma}
\noindent\emph{Proof.}\; We can decompose the gradients as follows:
\begin{equation}
\begin{split}
\mathbb{E} \left\| \nabla \mathbf{\Tilde{F}}(\mathbf{W}_k) \right\|_\mathbf{M}^2
&\overset{(a)}{=} \mathbb{E} \left\| \nabla \mathbf{\Tilde{F}}(\mathbf{W}_k) 
- \nabla \mathbf{\Tilde{F}}(\bm{u}_k\mathbf{1}_C^T) + \nabla \mathbf{\Tilde{F}}(\bm{u}_k\mathbf{1}_C^T) \right.\\
&\qquad \left.- \nabla \mathbf{\Tilde{F}}(\bm{u}_k) \mathbf{1}_C^T + \nabla \mathbf{\Tilde{F}}(\bm{u}_k) \mathbf{1}_C^T
- \overline{\nabla\mathbf{F}}(\mathbf{W}_k) \mathbf{1}_C^T + \overline{\nabla\mathbf{F}}(\mathbf{W}_k) \mathbf{1}_C^T \right\|_\mathbf{M}^2 \\
&\leq 4\mathbb{E} \left\| \nabla \mathbf{\Tilde{F}}(\mathbf{W}_k) 
- \nabla \mathbf{\Tilde{F}}(\bm{u}_k\mathbf{1}_C^T) \right\|_\mathbf{M}^2 
+ 4\mathbb{E} \left\| \nabla \mathbf{\Tilde{F}}(\bm{u}_k\mathbf{1}_C^T) - \nabla \mathbf{\Tilde{F}}(\bm{u}_k) \mathbf{1}_C^T \right\|_\mathbf{M}^2 \\
&\qquad + 4\mathbb{E} \left\| \nabla \mathbf{\Tilde{F}}(\mathbf{W}_k) \mathbf{1}_C^T - \overline{\nabla\mathbf{F}}(\mathbf{W}_k) \mathbf{1}_C^T \right\|_\mathbf{M}^2 
+ 4\mathbb{E} \left\| \overline{\nabla\mathbf{F}}(\mathbf{W}_k) \mathbf{1}_C^T \right\|_\mathbf{M}^2 \\
&\overset{(b)}{=} 4L^2 E_k + 4\mathbb{E} \left\| \nabla \mathbf{\Tilde{F}}(\bm{u}_k\mathbf{1}_C^T) - \nabla \mathbf{\Tilde{F}}(\bm{u}_k) \mathbf{1}_C^T \right\|_\mathbf{M}^2 + 4L^2 E_k + 4J_k \\
&\overset{(c)}{\leq} 8L^2 E_k + 4\kappa^2 + 4J_k.
\end{split}
\end{equation}
where we denote $\overline{\nabla\mathbf{F}}(\mathbf{W}_k) \triangleq \nabla\mathbf{\Tilde{F}}_k \bm{m}^T$ for simplicity in (a), (b) follows the definition of $J_k$ and $E_k$, and (c) is due to the assumption of non-IIDness in \eqref{kappa}.
\QEDB
\begin{lemma}\label{lemma-T1}
Denote $T_1(j,l,f)\!\triangleq\! 2\eta^2 \mathbb E \left[\left\| \sum\limits_{r=0}^{j-1} (Y_r-Q_r)(\mathbf{H}_2^{j-r} - \mathbf{M}) \!+\! (Y_{j,l}^{(1)}-Q_{j,l}^{(1)})(\mathbf{H}_1-\mathbf{M}) \!+\! (Y_{j,l,f}^{(2)}-Q_{j,l,f}^{(2)})(\mathbf{I}-\mathbf{M}) \right\|_{\mathbf{M}}^2 \right] $. Summing $T_1(j,l,f)$ over all iterations yields an upper bound as follows:
\begin{equation}
    \sum_{j=0}^{K/(\tau_1\tau_2)-1} \sum_{l=0}^{\tau_2-1} \sum_{f=1}^{\tau_1} T_1
    \leq 2\eta^2 K \left( \tau_1\tau_2 \frac{\zeta^{2\alpha}}{1-\zeta^{2\alpha}} + \frac{\tau_1\tau_2-1}{2} \right) \sigma^2.
\end{equation}
\end{lemma}
\noindent\emph{Proof.}\; According to the definition of $T_1(j,l,f)$, we have
\begin{equation}
    \begin{split}
    T_1(j,l,f) &= 2\eta^2 \left( \sum_{r=0}^{j-1} \mathbb E \left\| (Y_r-Q_r)(\mathbf{H}_2^{j-r}-\mathbf{M}) \right\|_\mathbf{M}^2
    + \mathbb E \left\| (Y_{j,l}^{(1)}-Q_{j,l}^{(1)})(\mathbf{H}_1-\mathbf{M}) \right\|_\mathbf{M}^2
    + \mathbb E \left\| (Y_{j,l,f}^{(2)}-Q_{j,l,f}^{(2)})(\mathbf{I}-\mathbf{M}) \right\|_\mathbf{M}^2\right) \\
    &= 2\eta^2 \left( \sum_{r=0}^{j-1} \mathbb E \left\| Y_r-Q_r \right\|_\mathbf{M}^2 \zeta^{2\alpha(j-r)}
    + \mathbb E \left\| Y_{j,l}^{(1)}-Q_{j,l}^{(1)} \right\|_\mathbf{M}^2
    + \mathbb E \left\| Y_{j,l,f}^{(2)}-Q_{j,l,f}^{(2)} \right\|_\mathbf{M}^2\right) \\
    &\overset{(a)}{\leq} 2\eta^2 \left(\sum_{r=0}^{j-1} \tau_1\tau_2 \sigma^2 \zeta^{2\alpha(j-r)}
    + l\tau_1 \sigma^2 + (f-1) \sigma^2 \right)\\
    &\leq 2\eta^2 \left( \tau_1\tau_2 \frac{\zeta^{2\alpha}}{1-\zeta^{2\alpha}}
    + l\tau_1 + f-1\right) \sigma^2,
    \label{eq-T-1}
    \end{split}
\end{equation}
where (a) is due to the following inequalities:
\begin{equation*}
    \mathbb{E} \left\| Y_{j,l}^{(1)} - Q_{j,l}^{(1)} \right\|_{\mathbf{M}}^2 \leq l\tau_1\sigma^2,\,
    \mathbb{E} \left\| Y_{j,l,f}^{(2)} - Q_{j,l,f}^{(2)} \right\|_{\mathbf{M}}^2 \leq (f-1)\sigma^2,\,
    \mathbb{E} \left\| Y_{r} - Q_{r} \right\|_{\mathbf{M}}^2 \leq \tau_1\tau_2\sigma^2.
\end{equation*}

We sum up $T_1(j,l,f)$ over $f=1,\dots,\tau_1$ and $l=0,\dots,\tau_2-1$, yielding the following results:
\begin{equation}
\begin{split}
    \sum_{l=0}^{\tau_2-1} \sum_{f=1}^{\tau_1} T_1
    & \leq 2\eta^2 \left[ \tau_1^2\tau_2^2 \frac{\zeta^{2\alpha}}{1-\zeta^{2\alpha}}
    + \frac{\tau_1^2\tau_2(\tau_2-1)}{2} + \frac{\tau_1\tau_2(\tau_1-1)}{2} \right] \sigma^2 \\
    & \leq 2\eta^2 \tau_1\tau_2 \left( \tau_1\tau_2 \frac{\zeta^{2\alpha}}{1-\zeta^{2\alpha}} + \frac{\tau_1\tau_2-1}{2} \right) \sigma^2.
    \label{eq-39}
\end{split}
\end{equation}
Finally, by summing up both sides of \eqref{eq-39} over $j=0,1,\dots,K/(\tau_1\tau_2)-1$, we have:
\begin{equation}
    \sum_{j=0}^{K/(\tau_1\tau_2)-1} \sum_{l=0}^{\tau_2-1} \sum_{f=1}^{\tau_1} T_1(j,l,f)
    \leq 2\eta^2 K \left( \tau_1\tau_2 \frac{\zeta^{2\alpha}}{1-\zeta^{2\alpha}} + \frac{\tau_1\tau_2-1}{2} \right) \sigma^2.
\end{equation}
\QEDB

\begin{lemma}\label{lemma-T2}
Denote $T_2(j,l,f)\triangleq 2\eta^2 \mathbb E \left[\left\| \sum_{r=0}^{j-1} Q_r(\mathbf{H}_2^{j-r} - \mathbf{M}) + Q_{j,l}^{(1)}(\mathbf{H}_1-\mathbf{M}) + Q_{j,l,f}^{(2)}(\mathbf{I}-\mathbf{M}) \right\|_{\mathbf{M}}^2 \right]$. Summing $T_2(j,l,f)$ over $k=1,2,\dots,K$ yields an upper bound as follows:
\begin{equation}
    \sum_{j=0}^{K/(\tau_1\tau_2)-1} \sum_{l=0}^{\tau_2-1} \sum_{f=1}^{\tau_1} T_2(j,l,f) \leq 2\tau_1\tau_2 \left( \tau_1\tau_2\Lambda + \frac{\tau_1\tau_2-1}{2} \frac{2-\zeta^\alpha}{1-\zeta^\alpha} \right) 
    \sum_{k=1}^K ( 8L^2E_k + 4\kappa^2 + 4J_k ),
\end{equation}
where $\Lambda \triangleq \frac{\zeta^{2\alpha}}{1-\zeta^{2\alpha}} + \frac{2\zeta^\alpha}{1-\zeta^\alpha} + \frac{\zeta^{2\alpha}}{(1-\zeta^\alpha)^2}$.
\end{lemma}
\noindent\emph{Proof.}\; According to the definition of $T_2(j,l,f)$, we have:
\begin{equation}
    \begin{split}
    T_2(j,l,f) &= 2\eta^2 \left( \sum_{r=0}^{j-1} \mathbb E \left\| Q_r (\mathbf{H}_2^{j-r}-\mathbf{M}) \right\|_\mathbf{M}^2
    + \mathbb E \left\| Q_{j,l}^{(1)}(\mathbf{H}_1-\mathbf{M}) \right\|_\mathbf{M}^2
    + \mathbb E \left\| Q_{j,l,f}^{(2)}(\mathbf{I}-\mathbf{M}) \right\|_\mathbf{M}^2\right) \\
    &\quad + \underbrace{2\eta^2 \sum_{r=0}^{j-1} \sum_{r^\prime=0,r^\prime\neq r}^{j-1} \mathbb E \left[ \underbrace{\Tr\left( (\mathbf{H}_2^{j-r}-\mathbf{M})Q_r^T Q_{r^\prime} (\mathbf{H}_2^{j-r^\prime}-\mathbf{M}) \right)}_{\text{Trace-1}} \right]}_{\text{Term-1}} \\
    &\quad + \underbrace{4\eta^2 \sum_{r=0}^{j-1} \mathbb E \left[ \Tr\left( (\mathbf{H}_1-\mathbf{M}){Q_{j,l}^{(1)}}^T Q_{r^\prime}(\mathbf{H}_2^{j-r^\prime}-\mathbf{M}) \right) \right]}_{\text{Term-2}} \\
    &\quad + \underbrace{4\eta^2 \sum_{r=0}^{j-1} \mathbb E \left[ \Tr\left( (\mathbf{I}-\mathbf{M}){Q_{j,l,f}^{(2)}}^T Q_r (\mathbf{H}_2^{j-r}-\mathbf{M}) \right) \right]}_{\text{Term-3}} \\
    &\quad + \underbrace{4\eta^2 \sum_{r=0}^{j-1} \mathbb E \left[ \Tr\left( (\mathbf{H}_1-\mathbf{M}) {Q_{j,l}^{(1)}}^T Q_{j,l,f}^{(2)} (\mathbf{I}-\mathbf{M}) \right) \right]}_{\text{Term-4}},
    \label{eq-T-2}
    \end{split}
\end{equation}
where the four traces in Terms 1-4 can be bounded similarly and thus we only bound the first trace, namely Trace-1, as follows:
\begin{equation}
    \begin{split}
    \text{Trace-1} &\leq \left\| (\mathbf{H}_2^{j-r}-\mathbf{M})Q_r \right\|_\mathbf{M} 
    \left\| Q_{r^\prime} (\mathbf{H}_2^{j-r^\prime}-\mathbf{M}) \right\|_\mathbf{M} \\
    &\leq \left\| \mathbf{H}_2^{j-r}-\mathbf{M} \right\|_\text{op} \left\| Q_r \right\|_\mathbf{M}
    \left\| Q_{r^\prime} \right\|_\mathbf{M} \left\| \mathbf{H}_2^{j-r^\prime}-\mathbf{M} \right\|_\text{op} \\
    &\leq \zeta^{(2j-r-r^\prime)\alpha} \left\| Q_r \right\|_\mathbf{M} \left\| Q_{r^\prime} \right\|_\mathbf{M} \\
    &\leq \frac{1}{2} \zeta^{(2j-r-r^\prime)\alpha} \left[ \left\| Q_r \right\|_\mathbf{M}^2 + \left\| Q_{r^\prime} \right\|_\mathbf{M}^2 \right].
    \end{split}
\end{equation}
As a result, the sum of four terms in the RHS of \eqref{eq-T-2} is bounded as follows:
\begin{equation}
    \begin{split}
    &\quad\quad \text{Term-1} + \text{Term-2} + \text{Term-3} + \text{Term-4} \\
    &\leq \eta^2 \sum_{r=0}^{j-1} \sum_{r^\prime=0,r^\prime\neq r}^{j-1} \zeta^{(2j-r-r^\prime)\alpha} 
    \left[\mathbb E \left\| Q_r \right\|_\mathbf{M}^2 + \mathbb E \left\| Q_r^\prime \right\|_\mathbf{M}^2 \right] \\
    & + 2\eta^2 \sum_{r=0}^{j-1} \zeta^{(j-r)\alpha} 
    \left[\mathbb E \left\| Q_{j,l}^{(1)} \right\|_\mathbf{M}^2 + \mathbb E \left\| Q_r \right\|_\mathbf{M}^2 \right]
    + 2\eta^2 \sum_{r=0}^{j-1} \zeta^{(j-r)\alpha} 
    \left[\mathbb E \left\| Q_{j,l,f}^{(2)} \right\|_\mathbf{M}^2 + \mathbb E \left\| Q_r \right\|_\mathbf{M}^2 \right]
    \end{split}
\end{equation}
\begin{equation*}
    \begin{split}
    & + 2\eta^2 \sum_{r=0}^{j-1} 
    \left[\mathbb E \left\| Q_{j,l}^{(1)} \right\|_\mathbf{M}^2 + \mathbb E \left\| Q_{j,l,f}^{(2)} \right\|_\mathbf{M}^2 \right]\\
    &\leq 2\eta^2 \sum_{r=0}^{j-1} \zeta^{(j-r)\alpha} \mathbb E \left\| Q_r \right\|_\mathbf{M}^2 \sum_{r^\prime=0,r^\prime\neq r}^{j-1} \zeta^{(j-r)\alpha}
    + 4\eta^2 \sum_{r=0}^{j-1} \zeta^{(j-r)\alpha} \mathbb E \left\| Q_l \right\|_\mathbf{M}^2 \\
    & + 2\eta^2 \mathbb E \left\| Q_{j,l}^{(1)} \right\|_\mathbf{M}^2 \sum_{r=0}^{j} \zeta^{(j-r)\alpha}  + 2\eta^2 \mathbb E \left\| Q_{j,l,f}^{(2)} \right\|_\mathbf{M}^2 \sum_{r=0}^{j} \zeta^{(j-r)\alpha}.
    \end{split}
\end{equation*}

We further observe that:
\begin{align}
    &\mathbb{E} \left\| Q_r \right \|_{\mathbf{M}}^2 \leq \tau_1\tau_2 \sum_{s=1}^{\tau_1\tau_2} \mathbb{E} \left\| \nabla \mathbf{\Tilde{F}} (\mathbf{W}_{r\tau_2\tau_1+s}) \right\|_{\mathbf{M}}^2, \label{eq-42}\\
    &\mathbb{E} \left\| Q_{j,l}^{(1)} \right \|_{\mathbf{M}}^2 \leq l\tau_1 \sum_{s=1}^{l\tau_1} \mathbb{E} \left\| \nabla \mathbf{\Tilde{F}} (\mathbf{W}_{j\tau_2\tau_1+s}) \right\|_{\mathbf{M}}^2, \label{eq-43} \\
    &\mathbb{E} \left\| Q_{j,l,f}^{(2)} \right \|_{\mathbf{M}}^2 \leq (f-1) \sum_{s=1}^{f-1} \mathbb{E} \left\| \nabla \mathbf{\Tilde{F}} (\mathbf{W}_{j\tau_2\tau_1+l\tau_1+s}) \right\|_{\mathbf{M}}^2. \label{eq-44}
\end{align}
By defining $\Lambda_{j-r} \triangleq \zeta^{2(j-r)\alpha} + 2\zeta^{(j-r)\alpha} + \frac{\zeta^{(j-r+1)\alpha}}{1-\zeta^\alpha}$, we have
\begin{equation}
    \sum_{j=0}^{K/\tau_1\tau_2-1}\sum_{r=0}^{j-1} \Lambda_{j-r} \leq \frac{K}{\tau_1\tau_2} \Lambda,
    \label{eq-lambda}
\end{equation}
where $\Lambda \triangleq \frac{\zeta^{2\alpha}}{1-\zeta^{2\alpha}} + \frac{2\zeta^\alpha}{1-\zeta^\alpha} + \frac{\zeta^{2\alpha}}{(1-\zeta^\alpha)^2}$.
Thus, $T_2(j,l,f)$ can be bounded as follows:
\begin{equation}
    \begin{split}
    T_2(j,l,f) &\leq 2\eta^2 \sum_{r=0}^{j-1}
    \left( \zeta^{2(j-r)\alpha} + 2\zeta^{(j-r)\alpha} + \frac{\zeta^{(j-r+1)\alpha}}{1-\zeta^\alpha} \right)
    \mathbb{E} \left\| Q_r \right\|_\mathbf{M}^2 \\
    &\quad+ 2\eta^2 \left( \frac{2-\zeta^\alpha}{1-\zeta^\alpha} \right) \mathbb{E} \left\| Q_{j,l}^{(1)} \right\|_\mathbf{M}^2
    + 2\eta^2 \left( \frac{2-\zeta^\alpha}{1-\zeta^\alpha} \right) \mathbb{E} \left\| Q_{j,l,f}^{(2)} \right\|_\mathbf{M}^2 \\
    &\overset{(a)}{=} 2\eta^2 \sum_{r=0}^{j-1} \Lambda_{j-r}
    \mathbb{E} \left\| \sum_{s=1}^{\tau_1\tau_2} \nabla \mathbf{\Tilde{F}}  (\mathbf{W}_{r\tau_1\tau_2+s}) \right\|_\mathbf{M}^2 \\
    &\quad+ 2\eta^2 \left( \frac{2-\zeta^\alpha}{1-\zeta^\alpha} \right) \mathbb{E} \left\|  \sum_{s=1}^{l\tau_1} \nabla \mathbf{\Tilde{F}} (\mathbf{W}_{j\tau_1\tau_2+s}) \right\|_\mathbf{M}^2
    + 2\eta^2 \left( \frac{2-\zeta^\alpha}{1-\zeta^\alpha} \right) \mathbb{E} \left\| \sum_{s=1}^{f-1} \nabla \mathbf{\Tilde{F}}  (\mathbf{W}_{j\tau_1\tau_2+l\tau_1+s}) \right\|_\mathbf{M}^2\\
    &\overset{(b)}{\leq} 2\eta^2 \tau_1 \tau_2 \Lambda
    \sum_{s=1}^{\tau_1\tau_2} \mathbb{E} \left\| \nabla \mathbf{\Tilde{F}}(\mathbf{W}_{r\tau_1\tau_2+s}) \right\|_\mathbf{M}^2 \\
    &\quad+ 2\eta^2 l\tau_1 \left( \frac{2-\zeta^\alpha}{1-\zeta^\alpha} \right) \sum_{s=1}^{l\tau_1} \mathbb{E} \left\| \nabla \mathbf{\Tilde{F}}(\mathbf{W}_{j\tau_1\tau_2+s}) \right\|_\mathbf{M}^2 
    + 2\eta^2 (f-1) \left( \frac{2-\zeta^\alpha}{1-\zeta^\alpha} \right) \sum_{s=1}^{f-1} \mathbb{E} \left\| \nabla \mathbf{\Tilde{F}}(\mathbf{W}_{j\tau_1\tau_2+l\tau_1+s}) \right\|_\mathbf{M}^2,
    \end{split}
\end{equation}
where (a) is due to \eqref{eq-42}-\eqref{eq-44} and (b) follows \eqref{eq-lambda}. By summing $T_2(j,l,f)$ over $f=1,\dots,\tau_1$ and $l=0,\dots,\tau_2-1$, we obtain:
\begin{equation}
    \begin{split}
    \sum_{l=0}^{\tau_2-1} \sum_{f=1}^{\tau_1} & T_2(j,l,f) 
    \leq 2\eta^2\tau_1^2\tau_2^2 \sum_{r=1}^{j-1} \Lambda_{j-r}
    \sum_{s=1}^{\tau_1\tau_2} \mathbb{E} \left\| \nabla \mathbf{\Tilde{F}}(\mathbf{W}_{r\tau_1\tau_2+s}) \right\|_{\mathbf{M}}^2 \\
    &+ 2\eta^2 \left( \frac{2-\zeta^\alpha}{1-\zeta^\alpha} \right) 
    \sum_{l=0}^{\tau_2-1} \sum_{f=1}^{\tau_1} 
    \left( l\tau_1 \sum_{s=1}^{l\tau_1} \mathbb{E} \left\| \nabla \mathbf{\Tilde{F}} (\mathbf{W}_{j\tau_1\tau_2+s}) \right\|_{\mathbf{M}}^2 
    + (f-1) \sum_{s=1}^{f-1} \mathbb{E} \left\| \nabla \mathbf{\Tilde{F}} (\mathbf{W}_{j\tau_1\tau_2+l\tau_1+s}) \right\|_{\mathbf{M}}^2 \right).
    \label{eq-47}
    \end{split}
\end{equation}
Notice that for any function $\beta(s)$, we have:
\begin{equation}
    \sum_{j=0}^{K/(\tau_1\tau_2)-1} \sum_{l=0}^{\tau_2-1} \sum_{f=1}^{\tau_1} \left(l\tau_1\sum_{s=j\tau_2\tau_1}^{j\tau_2\tau_1+l\tau_1} \beta(s)
    + (f-1) \sum_{s=j\tau_2\tau_1+l\tau_1+1}^{j\tau_2\tau_1+l\tau_1+f-1} \beta(s)\right)
    \leq \left( \frac{\tau_2(\tau_2-1)}{2}\tau_1^2 + \frac{\tau_1(\tau_1-1)}{2}\tau_2 \right)
    \sum_{k=1}^K \beta(k).
    \label{eq-beta}
\end{equation}
By further summing both sides of \eqref{eq-47} over $j=0,1,\dots,K/(\tau_1\tau_2)-1$ and applying \eqref{eq-beta}, we have:
\begin{equation}
    \begin{split}
    \sum_{j=0}^{K/(\tau_1\tau_2)-1} \sum_{l=0}^{\tau_2-1} \sum_{f=1}^{\tau_1} T_2(j,l,f)
    &\leq 2\eta^2\tau_1^2\tau_2^2 \Lambda
    \sum_{j=1}^{K/\tau_1\tau_2-1} \sum_{s=1}^{\tau_1\tau_2} \mathbb{E} \left\| \nabla \mathbf{\Tilde{F}} (\mathbf{W}_{r\tau_1\tau_2+s}) \right\|_\mathbf{M}^2 \\
    &\qquad+ 2\eta^2 \left( \frac{\tau_2(\tau_2-1)}{2}\tau_1^2 + \frac{\tau_1(\tau_1-1)}{2}\tau_2 \right) \left( \frac{2-\zeta^\alpha}{1-\zeta^\alpha} \right)
    \sum_{k=1}^K \mathbb{E} \left\| \nabla \mathbf{\Tilde{F}} (\mathbf{W}_{k}) \right\|_\mathbf{M}^2 \\
    &\overset{(c)}{\leq} 2\eta^2\tau_1^2\tau_2^2 \Lambda
    \sum_{k=1}^K \mathbb{E} \left\| \nabla \mathbf{\Tilde{F}} (\mathbf{W}_{k}) \right\|_\mathbf{M}^2 \\
    &\qquad+ 2\eta^2 \left( \frac{\tau_2(\tau_2-1)}{2}\tau_1^2 + \frac{\tau_1(\tau_1-1)}{2}\tau_2 \right) \left( \frac{2-\zeta^\alpha}{1-\zeta^\alpha} \right)
    \sum_{k=1}^K \mathbb{E} \left\| \nabla \mathbf{\Tilde{F}} (\mathbf{W}_{k}) \right\|_\mathbf{M}^2 \\
    &= 2\eta^2\tau_1\tau_2 \left( \tau_1\tau_2\Lambda + \frac{\tau_1\tau_2-1}{2} \frac{2-\zeta^\alpha}{1-\zeta^\alpha} \right)
    \sum_{k=1}^K \mathbb{E} \left\| \nabla \mathbf{\Tilde{F}} (\mathbf{W}_{k}) \right\|_\mathbf{M}^2 \\
    &\overset{(d)}{\leq} 2\tau_1\tau_2 \left( \tau_1\tau_2\Lambda + \frac{\tau_1\tau_2-1}{2} \frac{2-\zeta^\alpha}{1-\zeta^\alpha} \right) 
    \sum_{k=1}^K ( 8L^2E_k + 4\kappa^2 + 4J_k ),
    \end{split}
\end{equation}
where (c) follows \eqref{eq-beta} and (d) uses the results derived in Lemma \ref{lemma-global-gradient}.
\QEDB

\subsection{Proof of Lemma \ref{lemma-2}}\label{appendix-lemma-2}
First, we rewrite $\mathbf{W}_{k}(\mathbf{I}-\mathbf{M})$ as follows:
\begin{equation}
\begin{split}
    &\quad \mathbf{W}_{k}(\mathbf{I}-\mathbf{M}) \\
    &\overset{\text{(a)}}{=} \left( (\mathbf{W}_{k-1} -\eta \mathbf{G}_{k-1}) \mathbf{T}_{k-1} (\mathbf{I}-\mathbf{M}) \right)\\
    &\overset{\text{(b)}}{=} \mathbf{W}_{k-1}(\mathbf{I}-\mathbf{M}) \mathbf{T}_{k-1} - \eta \mathbf{G}_{k-1} (\mathbf{T}_{k-1}-\mathbf{M}) \\
    &\overset{\text{(c)}}{=} \left[ (\mathbf{W}_{k-2} - \eta \mathbf{G}_{k-2}) \mathbf{T}_{k-2} (\mathbf{I}-\mathbf{M}) \right]\mathbf{T}_{k-1} - \eta \mathbf{G}_{k-1} (\mathbf{T}_{k-1}-\mathbf{M}) \\
    &\overset{\text{(d)}}{=} \mathbf{W}_{k-2}(\mathbf{I}-\mathbf{M}) \mathbf{T}_{k-2} \mathbf{T}_{k-1} - \eta \mathbf{G}_{k-2} (\mathbf{T}_{k-2}\mathbf{T}_{k-1}-\mathbf{M}) \eta \mathbf{G}_{k-1} (\mathbf{T}_{k-1}-\mathbf{M}),
\end{split}
\end{equation}
where (a) and (c) follow \eqref{eq-relation}, while (b) and (d) follow Lemma \ref{lemma-3}.
By mathematical induction, we obtain:
\begin{equation}
    \mathbf{W}_{k}(\mathbf{I}-\mathbf{M}) 
    = \mathbf{W}_0(\mathbf{I}-\mathbf{M}) \prod_{l=0}^{k-1}\mathbf{T}_{l} - \eta \sum_{s=0}^{k-1}\mathbf{G}_{s}\left( \prod_{l=s}^{k-1}\mathbf{T}_{l} - \mathbf{M} \right)\\
    \overset{\text{(e)}}{=} \eta \sum_{s=0}^{k-1}\mathbf{G}_{s}\left( \prod_{l=s}^{k-1}\mathbf{T}_{l} - \mathbf{M} \right),
    \label{eq-W-k}
\end{equation}
where (e) holds since $\mathbf{W}_0 = [\bm{w}_0,\dots,\bm{w}_0]$, i.e., $\mathbf{W}_0(\mathbf{I}-\mathbf{M})=0$.

Denote $\Phi_{s,k-1} \triangleq \prod_{l=s}^{k-1}\mathbf{T}_{l}$, which can be expressed as:
\begin{equation}
    \Phi_{s,k-1} = \left\{
        \begin{array}{ll}
        \mathbf{I}, & \text{if } j\tau_2\tau_1+l\tau_1 < s < j\tau_2\tau_1+l\tau_1+f, \\
        \mathbf{H}_1, & \text{if } j\tau_2\tau_1 < s \leq j\tau_2\tau_1+l\tau_1, \\
        \mathbf{H}_2, & \text{if } (j-1)\tau_2\tau_1+l\tau_1 < s \leq j\tau_2\tau_1, \\
        \dots & \\
        \mathbf{H}_2^j, & \text{if } 1 \leq s \leq \tau_2\tau_1.
        \end{array}
        \right.
    \label{eq-phi}
\end{equation}
With the results in \eqref{eq-W-k} and \eqref{eq-phi}, we derive an upper bound for $\mathbb E \left[\left\| \mathbf{W}_k (\mathbf{I}-\mathbf{M}) \right\|_{\mathbf{M}}^2 \right]$ as follows:
\begin{equation*}
\begin{split}
    \mathbb E \left[\left\| \mathbf{W}_k(\mathbf{I}-\mathbf{M}) \right\|_{\mathbf{M}}^2 \right]
    &= \eta^2 \mathbb E \left[\left\| \sum_{r=0}^{j-1} Y_r(\mathbf{H}_2^{j-r} - \mathbf{M}) + Y_{j,l}^{(1)}(\mathbf{H}_1-\mathbf{M}) + Y_{j,l,f}^{(2)}(\mathbf{I}-\mathbf{M}) \right\|_{\mathbf{M}}^2 \right]\\
    &\leq \underbrace{2\eta^2 \mathbb E \left[\left\| \sum_{r=0}^{j-1} (Y_r-Q_r)(\mathbf{H}_2^{j-r} - \mathbf{M}) + (Y_{j,l}^{(1)}-Q_{j,l}^{(1)})(\mathbf{H}_1-\mathbf{M}) + (Y_{j,l,f}^{(2)}-Q_{j,l,f}^{(2)})(\mathbf{I}-\mathbf{M}) \right\|_{\mathbf{M}}^2 \right]}_{T_1(j,l,f)} \\
\end{split}
\end{equation*}
\begin{equation}
\begin{split}
    &+ \underbrace{2\eta^2 \mathbb E \left[\left\| \sum_{r=0}^{j-1} Q_r(\mathbf{H}_2^{j-r} - \mathbf{M}) + Q_{j,l}^{(1)}(\mathbf{H}_1-\mathbf{M}) + Q_{j,l,f}^{(2)}(\mathbf{I}-\mathbf{M}) \right\|_{\mathbf{M}}^2 \right]}_{T_2(j,l,f)}.
    \label{eq-E}
\end{split}
\end{equation}

We first sum up both sides of \eqref{eq-E} over all iterations and further apply the bounds of $T_1(j,l,f)$ and $T_2(j,l,f)$ derived in Lemma \ref{lemma-T1} and Lemma \ref{lemma-T2} respectively, yielding:
\begin{equation}
    \begin{split}
    \frac{1}{K} \sum_{k=1}^{K} E_k &= \frac{1}{K} \sum_{k=1}^{K} \mathbb E \left\| \mathbf{W}_k(\mathbf{I}-\mathbf{M}) \right\|_\mathbf{M}^2 \\
    &\leq \frac{1}{K} \bigg[ 2\eta^2 K \left( \tau_1\tau_2 \frac{\zeta^{2\alpha}}{1-\zeta^{2\alpha}} + \frac{\tau_1\tau_2-1}{2} \right) \sigma^2 \\
    &\qquad + 2 \tau_1\tau_2 \left( \tau_1\tau_2\Lambda + \frac{\tau_1\tau_2-1}{2} \frac{2-\zeta^\alpha}{1-\zeta^\alpha} \right) 
    \sum_{k=1}^K ( 8L^2E_k + 4\kappa^2 + 4J_k ) \bigg]\\
    &\overset{(f)}{=} 2\eta^2  V_1^\prime \sigma^2 + 16\eta^2L^2 V_3 \left(\frac{1}{K} \sum_{k=1}^{K} E_k \right) + 8\eta^2 V_3 \kappa^2 +  8\eta^2 V_3 \left( \frac{1}{K} \sum_{k=1}^{K} J_k \right),
    \end{split}
\end{equation}
where we denote $V_1^\prime \triangleq \tau_1\tau_2 \frac{\zeta^{2\alpha}}{1-\zeta^{2\alpha}} + \frac{\tau_1\tau_2-1}{2} $ and $V_3 \triangleq \tau_1\tau_2 \left(\tau_1\tau_2\Lambda + \frac{\tau_1\tau_2-1}{2} \frac{2-\zeta^\alpha}{1-\zeta^\alpha} \right)$ in (f). When $1- 16\eta^2L^2V_3>0$, we further obtain the following inequality:
\begin{equation}
    (1- 16\eta^2L^2V_3) \frac{1}{K} \sum_{k=1}^{K} E_k \leq 
    2\eta^2  V_1^\prime \sigma^2 + 8\eta^2 V_3 \kappa^2 +  8\eta^2 V_3 \left( \frac{1}{K} \sum_{k=1}^{K} J_k \right).
    \label{eq:E_kV_3}
\end{equation}
We complete the proof by dividing both sides of \eqref{eq:E_kV_3} by $V_1=V_1^\prime/(1-16\eta^2L^2V_3)$ and $V_2=V_3/(1-16\eta^2L^2V_3)$.
\QEDB
\subsection{Proof of Theorem \ref{thm-1}}\label{appendix-thm}
First, we sum up \eqref{eq-one-step} in Lemma \ref{lemma-one-step} for $k=1,\dots,K$, and divide both sides by $K$ as follows:
\begin{equation}
    \begin{split}
    \frac{1}{K} \left( \mathbb{E}[F(\bm{u}_{k+1})]- \mathbb{E}F(\bm{u}_1) \right)
    & \leq -\frac{\eta}{2K} \sum_{k=1}^{K} \mathbb{E} \left\| \nabla F(\bm{u}_k) \right\|_2^2 +  \frac{\eta^2L}{2} \sum_{i\in\mathcal{C}} m_i^2 \sigma^2 \\
    & + \frac{\eta L^2}{2} \frac{1}{K} \sum_{k=1}^{K} \mathbb{E} \left\| \mathbf{W}_k(\mathbf{I}-\mathbf{M}) \right\|_{\mathbf{M}}^2
    - \frac{\eta}{2} (1 - \eta L) \frac{1}{K} \sum_{k=1}^{K} J_k.
    \end{split}
\end{equation}
By rearranging terms, multiplying each of them by $\frac{2}{\eta}$, we arrive at the following expression:
\begin{equation}
    \begin{split}
    &\quad \frac{1}{K} \sum_{k=1}^{K} \mathbb{E} \left\| \nabla F(\bm{u}_k) \right\|_2^2 \\
    &\overset{(a)}{\leq} \frac{2\left[\mathbb{E}[F(\bm{u}_{k+1})]- \mathbb{E}F(\bm{u}_1)\right]}{\eta K} + \eta L \sum_{i\in\mathcal{C}} m_i^2 \sigma^2 \\
    &\qquad- (1 - \eta L) \frac{1}{K} \sum_{k=1}^{K} J_k + L^2 \left(2\eta^2 V_1 \sigma^2 + 8\eta^2 V_2 \kappa^2 + 8\eta^2 V_2 \frac{1}{K} \sum_{k=1}^{K} J_k \right) \\
    &\overset{(b)}{=} \frac{2\Delta}{\eta K} + \eta L \sum_{i\in\mathcal{C}} m_i^2 \sigma^2 + 2\eta^2L^2 V_1 \sigma^2 + 8\eta^2 L^2 V_2 \kappa^2
     - \left( 1 - \eta L - 8\eta^2L^2 V_2 \right) \frac{1}{K} \sum_{k=1}^{K} J_k,
    \end{split}
\end{equation}
where (a) follows Lemma \ref{lemma-2}, and (b) denotes $\Delta\triangleq \mathbb{E}\left[F(\bm{u}_1)\right]- \mathbb{E}\left[F(\bm{u}^*)\right] \geq \mathbb{E}\left[F(\bm{u}_1)\right]- \mathbb{E}\left[F(\bm{u}_{K+1})\right]$. As a result, if $\eta$ satisfies 
\begin{equation}
    1-\eta L-8\eta^2L^2 V_2 \geq 0, 1-16\eta^2L^2V_3 >0,
\end{equation}
we have:
\begin{equation}
\begin{split}
    \frac{1}{K} \sum_{k=1}^{K} \mathbb{E} \left\| \nabla F(\bm{u}_k) \right\|_2^2
    \leq \frac{2\Delta}{\eta K} + \eta L \sum_{i\in\mathcal{C}} m_i^2 \sigma^2 + 2\eta^2L^2 V_1 \sigma^2 + 8\eta^2 L^2 V_2 \kappa^2,
\end{split}
\end{equation}
which completes the proof.
\QEDB



\begin{thebibliography}{99}
\bibitem{IoT} K. L. Lueth, “State of the IoT 2020: 12 billion IoT connections, surpassing non-IoT for the first time.” Nov. 2020. [Online]. Available: \url{https://iot-analytics.com/state-of-the-iot-2020-12-billion-iot-connections-surpassing-non-iot-for-the-first-time/}

\bibitem{meneghello2019iot} F. Meneghello, M. Calore, D. Zucchetto, M. Polese, and A. Zanella, “IoT: Internet of Threats? {A} survey of practical security vulnerabilities in real IoT devices,” \emph{IEEE Internet Things J.}, vol. 6, no. 5, pp. 8182–8201, Oct. 2019.

\bibitem{mcmahan2017communication} B. McMahan, E. Moore, D. Ramage, S. Hampson, and B. A. y Arcas, “Communication-efficient learning of deep networks from decentralized data,” in \emph{Proc. Int. Conf. Artif. Intell. Statist. (AISTATS)}, Ft. Lauderdale, FL, USA, Apr. 2017.

\bibitem{li2020federated} T. Li, A. K. Sahu, A. Talwalkar, and V. Smith, “Federated learning: Challenges, methods, and future directions,” \emph{IEEE Signal Process. Mag.}, vol. 37, no. 3, pp. 50–60, May 2020.

\bibitem{shi2020communication} Y. Shi, K. Yang, T. Jiang, J. Zhang, and K. B. Letaief, “Communication-efficient edge AI: Algorithms and systems,” \emph{IEEE Commun. Surveys Tuts.}, vol. 22, no. 4, pp. 2167–2191, 4th Quart. 2020.

\bibitem{mao2017survey} Y. Mao, C. You, J. Zhang, K. Huang, and K. B. Letaief, “A survey on mobile edge computing: The communication perspective,” \emph{IEEE Commun. Surveys Tuts.}, vol. 19, no. 4, pp. 2322-2358, 4th Quart. 2017.

\bibitem{lim2020federated} W. Y. B. Lim \textit{et al.}, “Federated learning in mobile edge networks: A comprehensive survey,” \emph{IEEE Commun. Surveys Tuts.}, vol. 22, no. 3, pp. 2031–2063, 3rd Quart., 2020.

\bibitem{wang2019adaptive} S. Wang \textit{et al.}, “Adaptive federated learning in resource constrained edge computing systems,” \emph{IEEE J. Sel. Areas Commun.}, vol. 37, no. 6, pp. 1205–1221, Mar. 2019.

\bibitem{nishio2019client} T. Nishio and R. Yonetani, “Client selection for federated learning with heterogeneous resources in mobile edge,” in \emph{Proc. IEEE Int. Conf. Commun. (ICC)}, Shanghai, China, May 2019.

\bibitem{mills2019communication} J. Mills, J. Hu, and G. Min, “Communication-efficient federated learning for wireless edge intelligence in IoT,” \emph{IEEE Internet Things J.}, vol. 7, no. 7, pp. 5986–5994, Jul. 2020.

\bibitem{amiri2020federated} M. M. Amiri and D. G{\"u}nd{\"u}z, “Federated learning over wireless fading channels,” \emph{IEEE Trans. Wireless Commun.}, vol. 19, no. 5, pp. 3546–3557, Feb. 2020.

\bibitem{liu2020client} L. Liu, J. Zhang, S. Song, and K. B. Letaief, “Client-edge-cloud hierarchical federated learning,” in \emph{Proc. IEEE Int. Conf. Commun.(ICC)}, Dublin, Ireland, Jun. 2020.

\bibitem{wang2020local} J. Wang, S. Wang, R.-R. Chen, and M. Ji, “Local averaging helps: Hierarchical federated learning and convergence analysis.” [Online]. Available: \url{https://arxiv.org/pdf/2010.12998.pdf}

\bibitem{saha2020fogfl} R. Saha, S. Misra, and P. K. Deb, “FogFL: Fog assisted federated learning for resource-constrained IoT devices,” \emph{IEEE Internet Things J.}, vol. 8, no. 10, pp. 8456-8463, May 2021.

\bibitem{castiglia2020multi} T. Castiglia, A. Das, and S. Patterson, “Multi-level local {SGD:} Distributed {SGD} for heterogeneous hierarchical networks,” in \emph{Proc. Int. Conf. Learn. Repr. (ICLR)}, Virtual Event, May 2021.

\bibitem{elsasser2002diffusion} R. Els{\"a}sser, B. Monien, and R. Preis, “Diffusion schemes for load balancing on heterogeneous networks,” \emph{Theory Comput. Syst.}, vol. 35, no. 3, pp. 305–320, Jun. 2002.

\bibitem{tang2018d} H. Tang, X. Lian, M. Yan, C. Zhang, and J. Liu, “$ D^{2}$: Decentralized training over decentralized data,” in \emph{Proc. Int. Conf. Mach. Learn. (ICML)}, Stockholm, Sweden, Jul. 2018.


\bibitem{parallelsgd} L. Bottou, F. E. Curtis, and J. Nocedal, “Optimization methods for large-scale machine learning,” \emph{SIAM Rev.}, vol. 60, no. 2, pp. 223-311, Aug. 2018.

\bibitem{wang2020federated} H. Wang, M. Yurochkin, Y. Sun, D. S. Papailiopoulos, and Y. Khazaeni, “Federated learning with matched averaging,” in \emph{Proc. Int. Conf. Learn. Repr. (ICLR)}, Addis Ababa, Ethiopia, Apr. 2020.

\bibitem{shi2020joint} W. Shi, S. Zhou, Z. Niu, M. Jiang, and L. Geng, “Joint device scheduling and resource allocation for latency constrained wireless federated learning,” \emph{IEEE Trans. Wireless Commun.}, vol. 20, no. 1, pp. 453-467, Jan. 2021.

\bibitem{hu2020coedge} L. Hu, G. Sun, and Y. Ren, “CoEdge: Exploiting the edge-cloud collaboration for faster deep learning,” \emph{{IEEE} Access}, vol. 8, pp. 100533-100541, Jul. 2020.

\end{thebibliography}
\end{document}